\documentclass[journal]{IEEEtran}
\usepackage{tabularx}
\usepackage{lmodern}
\usepackage{float}
\usepackage{textalpha}
\usepackage{cuted}
\usepackage{fancyvrb}
\usepackage{listings}
\usepackage{booktabs}
\DeclareUnicodeCharacter{2502}{|}
\DeclareUnicodeCharacter{2009}{\,}
\usepackage{listings}
\usepackage{amsmath,amsfonts}
\usepackage{pgfplots}
\usepackage{pgfplotstable}
\pgfplotsset{compat=1.18}
\usepackage{newtxtext}
\usepackage{tabularx}
\usepackage{booktabs}
\usepackage{graphicx}
\usepackage{tikz}
\usetikzlibrary{positioning,arrows.meta}
\usepackage{cite}
\usepackage{hyperref}
\usepackage[compatibility=false]{caption}
\usepackage{array}
\newcolumntype{L}[1]{>{\raggedright\arraybackslash}p{#1}}
\newcolumntype{C}[1]{>{\centering\arraybackslash}p{#1}}
\usepackage{subfig}
\usepackage{textalpha}
\usepackage{multirow}
\usepackage{setspace}
\usepackage{enumitem}
\usepackage{siunitx}
\usepackage{tikz}
\usetikzlibrary{shadows}
\begin{document}

\title{Building a Robust Risk-Based Access Control System to Combat Ransomware's Capability to Encrypt}

\author{
    \IEEEauthorblockN{Kenan Begovic}
    \IEEEauthorblockA{
        College of Engineering\\
        Qatar University\\
        Doha, Qatar\\
        Email: kb2000115@qu.edu.qa\\
    } 
\and
    \IEEEauthorblockN{Abdulaziz Al-Ali}
    \IEEEauthorblockA{
        College of Engineering\\
        Qatar University\\
        Doha, Qatar\\
    }
\and
    \IEEEauthorblockN{Qutaibah Malluhi}
    \IEEEauthorblockA{
        College of Engineering\\
        Qatar University\\
        Doha, Qatar\\
    }
}

\maketitle
\begin{abstract}
Ransomware’s core capability—unauthorized encryption—demands controls that identify and block malicious cryptographic activity without disrupting legitimate use. We present a probabilistic, risk-based access control architecture that couples machine-learning inference with mandatory access control to regulate encryption on Linux in real time. The system builds a specialized dataset from the native \texttt{ftrace} framework using the \texttt{function\_graph} tracer, yielding high-resolution kernel-function execution traces augmented with resource and I/O counters. These traces support both a supervised classifier and interpretable rules that drive an SELinux policy via lightweight booleans, enabling context-sensitive permit/deny decisions at the moment encryption begins. Compared to approaches centered on sandboxing, hypervisor introspection, or coarse system-call telemetry, the function-level tracing we adopt provides finer behavioral granularity than syscall-only telemetry while avoiding the virtualization/VMI overhead of sandbox-based approaches. Our current user-space prototype has a non-trivial footprint under burst I/O; we quantify it and recognize that a production kernel-space solution should aim to address this. We detail dataset construction, model training and rule extraction, and the run-time integration that gates file writes for suspect encryption while preserving benign cryptographic workflows. During evaluation, the two-layer composition retains model-level detection quality while delivering rule-like responsiveness; we also quantify operational footprint and outline engineering steps to reduce CPU and memory overhead for enterprise deployment. The result is a practical path from behavioral tracing and learning to enforceable, explainable, and risk-proportionate encryption control on production Linux systems.
\end{abstract}


\section{Introduction}

The escalating ransomware threat has underscored the need for robust, adaptable access control mechanisms capable of thwarting unauthorized encryption activities in real-time~\cite{mcintosh_ransomware_2025}. Ransomware, which maliciously encrypts valuable data to demand ransom, continues to evolve in sophistication, leveraging advanced cryptographic techniques and exploiting gaps in conventional security protocols. This paper responds to this threat by proposing an integrated framework that leverages risk-based access control principles to dynamically regulate encryption permissions, specifically within environments hardened by SELinux. Traditional access control models, although effective in static contexts, often lack the probabilistic adaptability required to meet the demands of modern cybersecurity\cite{srivastava2020machine}. Our objective is to build a solution capable of identifying and mitigating unauthorized encryption activities while preserving legitimate cryptographic operations, through the use of detectors built on machine learning to detect encryption and mandatory access control (MAC) that blocks unauthorized activities. This research is driven by the search for answers to questions about how to build specialized datasets to train machine learning (ML) models and what learning techniques and algorithms are applicable to the problem. We seek to explore how extracted rules from the ML models in our probabilistic access control system (ACS) perform compared to using the model itself. Also, how would such ACS be integrated into the operating system, and how effective would it be in real-life scenarios?
 
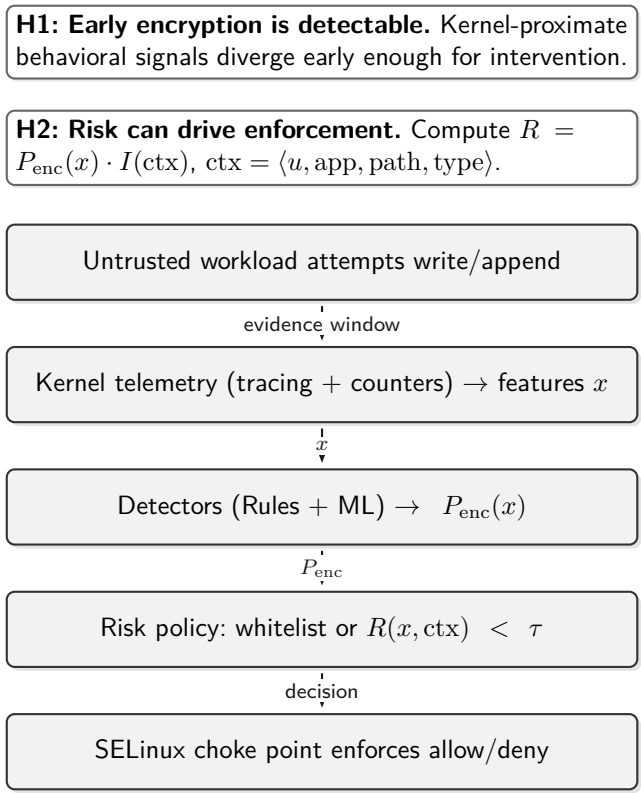
\begin{figure}[t]
\centering
\begin{tikzpicture}[
  font=\sffamily,
  >=Stealth,
  node distance=6mm,
  process/.style={
    rectangle, draw=black!80, thick, fill=gray!10,
    rounded corners=3pt, drop shadow={opacity=0.22, shadow xshift=1.5pt, shadow yshift=-1.5pt},
    minimum height=10mm, text width=0.92\columnwidth, align=center
  },
  hypo/.style={
    rectangle, draw=black!60, thick, fill=white,
    rounded corners=3pt, drop shadow={opacity=0.18, shadow xshift=1.2pt, shadow yshift=-1.2pt},
    text width=0.92\columnwidth, align=left
  },
  arrow/.style={->, thick, black!90, shorten >=2pt, shorten <=2pt},
  labeltext/.style={font=\sffamily\footnotesize, fill=white, inner sep=1.5pt}
]

\node (h1) [hypo] {\textbf{H1: Early encryption is detectable.}
Kernel-proximate behavioral signals diverge early enough for intervention.};

\node (h2) [hypo, below=4mm of h1] {\textbf{H2: Risk can drive enforcement.}
Compute $R=P_{\mathrm{enc}}(x)\cdot I(\mathrm{ctx})$, $\mathrm{ctx}=\langle u,\mathrm{app},\mathrm{path},\mathrm{type}\rangle$.};

\node (workload)  [process, below=5mm of h2] {Untrusted workload attempts write/append};
\node (telemetry) [process, below=of workload] {Kernel telemetry (tracing + counters) $\rightarrow$ features $x$};
\node (detectors) [process, below=of telemetry] {Detectors (Rules + ML) $\rightarrow P_{\mathrm{enc}}(x)$};
\node (risk)      [process, below=of detectors] {Risk policy: whitelist or $R(x,\mathrm{ctx})<\tau$};
\node (enforce)   [process, below=of risk] {SELinux choke point enforces allow/deny};

\draw[arrow] (workload.south) -- node[labeltext] {evidence window} (telemetry.north);
\draw[arrow] (telemetry.south) -- node[labeltext] {$x$} (detectors.north);
\draw[arrow] (detectors.south) -- node[labeltext] {$P_{\mathrm{enc}}$} (risk.north);
\draw[arrow] (risk.south) -- node[labeltext] {decision} (enforce.north);

\end{tikzpicture}
\caption{Compact hypothesis view: early behavioral evidence supports probabilistic detection, and a context-weighted risk policy drives OS-native enforcement through SELinux.}
\label{fig:hypothesis-compact}
\end{figure}

To address this gap, we took promising results from our initial investigation into the detection of encryption \cite{begovic2025exploiting} using ML and envisioned a hypothesis presented in Figure~\ref{fig:hypothesis-compact}. Our research employs a probabilistic access control system informed by a unique ML approach that can dynamically predict and detect encryption-related activities based on behavioral patterns observed within system traces. By utilizing \texttt{ftrace}'s \texttt{function\_graph} tracer, we gain a comprehensive view of kernel functions involved in critical processes, capturing intricate call relationships that provide valuable insights into system activities. This approach not only enables more precise monitoring but also facilitates real-time adaptation to emerging threats by identifying suspicious cryptographic patterns as they develop. Unlike prior research, which predominantly relies on extended Berkeley Packet Filter (eBPF), hypervisor tracing, or sandboxed behavioral monitoring, our framework uniquely leverages the above-mentioned framework tracer to collect low-level function execution data. This provides unprecedented granularity in modeling cryptographic behaviors and supports more accurate ML-based policy inference.

Our framework represents a significant advancement in access control for ransomware detection, offering both the flexibility of probabilistic risk-based models and the enhanced visibility provided by real-time tracing of system-level functions. Through the dynamic rule-extraction process from our trained ML model, the system can proactively modify encryption permissions, contributing to a more resilient defense against unauthorized access and cryptographic misuse. This integration is unique in its application of \texttt{ftrace} in machine learning, where function-level tracing is systematically analyzed to inform access control policies that are highly responsive to cryptographic behaviors, thus addressing a critical need for adaptability in high-security environments.

\vspace{0.25em}
\noindent The following summarizes our contributions:
\begin{itemize}
  \item \textbf{Dual-layer, risk-based enforcement.} We design and implement a two-layer detector that combines (i)~a rule layer (E\textsubscript{RULE}) and (ii)~a model layer (E\textsubscript{ML}) to gate encryption actions. Both layers consume the \emph{same 36-feature} vector (resource counters, two graph metrics—betweenness and clustering—and 27 function-level signals), enabling consistent early blocking with different cost/latency profiles.
  \item \textbf{Function-level tracing for security policy.} We operationalize Linux \texttt{ftrace}/\texttt{function\_graph} to capture kernel function activity and derive graph-structured features that characterize cryptographic behavior. To our knowledge, this is the first work to systematically translate function-level traces into ML-informed access-control decisions.
  \item \textbf{Feature selection pipelines and elbow analyses.} We present two complementary selection procedures from the dataset with 113 engineered features: (P1) permutation-importance with an elbow rule (which identifies a strict knee at \(k=2\) for a minimal shallow tree), and (P2) a \(\chi^2\)+utility wrapper that maximizes a defender-centric payoff. Although P1 selects two features at the elbow, we expand E\textsubscript{RULE} to the full 36-feature set to improve robustness and reduce latency spikes in practice.
  \item \textbf{Policy integration with SELinux using Common Intermediate Language (CIL).} We provide a minimal, auditable SELinux CIL module that introduces \texttt{encryption\_t} and two \emph{Boolean hooks} (\texttt{rule\_block}, \texttt{ml\_block}). User space daemons flip these toggles to allow or deny \texttt{write/append} on labeled file types. The policy supports (a)~\emph{application} whitelisting (\texttt{crypto\_exec\_t}), (b)~\emph{user} whitelisting, and (c)~\emph{path/type} scoping (e.g., \texttt{user\_home\_t}).
  \item \textbf{Principled whitelisting semantics.} We formalize a whitelist model in which specific users may run selected encryptors within declared file contexts (e.g., user1+OpenSSL under \texttt{\$HOME/**}), while all other combinations default to restrictive handling when either detector asserts risk.
  \item \textbf{Early-block strategy.} We target first tens of milliseconds / few–tens of kilobytes by sampling at sub-second cadence and by prioritizing short, high-signal features (graph metrics + hot kernel paths). Our logs report both \emph{decision latency} and \emph{bytes to the first block} as primary outcome metrics.
  \item \textbf{Evaluation and baselines.} We evaluate on a mixed benign corpus, policy-constrained crypto tools (OpenSSL, GnuPG), and Linux ransomware scenarios, and compare to two families of defenses announced in our baselines: an eBPF-based detector (Brodzik \emph{et al.}) and a Cuckoo v3.2 + GBDT pipeline.
  \item \textbf{Operational lessons.} We document practical pitfalls—CIL-only loading on Ubuntu 24.04, JSON/YAML-tolerant key ingestion for trace symbol lists, and resource behavior under load (RAM residency \(\sim\)50\%, frequent CPU spikes)—and the resulting engineering choices (using the full 36 features in E\textsubscript{RULE}, consolidating the collector).
  \item \textbf{Reproducibility scope.} We release relevant code for automation to reproduce the measurements (workload harnesses, report generation, SELinux modules, and daemon scaffolding) and the \emph{raw, non-processed dataset}. To prevent misuse, trained models and production rule sets are withheld; nevertheless, our artifacts deterministically regenerate the tables and figures from new runs.
\end{itemize}

\paragraph{Threat model and enforcement boundary}
This work targets the dominant operational ransomware setting in which an adversary achieves arbitrary \emph{user-space} code execution (e.g., via a dropper) and attempts to perform bulk file encryption over user and shared-data paths. We assume SELinux runs in \emph{enforcing} mode and that the attacker does not possess SELinux-administrative capabilities (e.g., cannot load policy modules, cannot relabel protected types, and cannot manipulate SELinux booleans via \texttt{setsebool}). Under these assumptions, the key enforcement surface is the SELinux authorization point: write/append permissions to protected file types are mediated by the SELinux LSM and can be revoked at run time via Boolean-guarded policy.

Our prototype’s scope is intentionally constrained: it is a proof-of-concept to validate (i) detector correctness under realistic benign/crypto/ransomware workloads and (ii) reliable end-to-end enforcement by integrating detector verdicts into SELinux policy hooks. We therefore treat kernel compromise, SELinux disablement/permissive mode, and arbitrary policy reload by a privileged adversary as out of scope; these represent a stronger attacker that defeats most host-based controls and requires orthogonal hardening (secure boot/lockdown, policy immutability, and attestation), discussed as deployment directions.

Enforcement applies to processes confined to the designated subject domain (\texttt{encryption\_t}) and to labeled assets (e.g., \texttt{user\_home\_t} / scoped file types). Consequently, the operational question “what runs in \texttt{encryption\_t}?” is part of the deployment posture, not a user action: in our experiments, we confine the target workload under \texttt{encryption\_t} for repeatability, while a production posture relies on policy-driven domain transitions (executable labeling and \texttt{type\_transition}) to ensure that untrusted processes are covered without cooperative wrappers.

To mitigate the risk of intentional Denial of Service (DoS) attacks—where an adversary might artificially induce ransomware-like I/O patterns in a benign critical process to provoke a system block—our enforcement boundary is strictly domain-isolated. Critical system daemons (e.g., database or backup services) do not run in the \texttt{encryption\_t} domain, meaning our ML and rule booleans (rule\_block / ml\_block) have no authority to revoke their write privileges. Enforcement is exclusively applied to untrusted processes that have been transitioned into the \texttt{encryption\_t} domain. Even if an attacker induces a false positive in a confined process, the resulting write/append denial is localized solely to that process's specific context, preventing a system-wide DoS.

The remainder of this paper is structured as follows. Section II reviews related work in access control models and encryption detection, highlighting both traditional and probabilistic approaches. Section III details our methodology, encompassing dataset construction, machine learning model training, feature selection, rule extraction, and SELinux integration. \label{sec:dataset} Section IV presents our experimental results and evaluates the framework's effectiveness in detecting ransomware encryption patterns. Section V discusses the broader implications of this work, including its potential for wider adoption in risk-based access control systems. Finally, Section VI concludes with a summary and suggests avenues for future research to further enhance resilience against evolving ransomware threats.

\section{Related Work}
\noindent\textbf{Comparative lens.}
Across prior work, ransomware defenses differ primarily along five axes: \emph{(i)} telemetry locus (user space, kernel, hypervisor, or storage device), \emph{(ii)} enforcement locus (alerting vs. inline access mediation), \emph{(iii)} reaction latency (early-block vs. post-hoc), \emph{(iv)} explainability at decision time (rules/signatures vs. opaque models), and \emph{(v)} deployability (privilege and infrastructure requirements). We use this lens below to clarify how existing approaches trade visibility, operational complexity, and enforcement strength.

\subsection{Access Control Models and Risk-Based Approaches}
Access control models have evolved significantly, from traditional frameworks such as Discretionary Access Control (DAC) and Mandatory Access Control (MAC), to more dynamic paradigms like Role-Based Access Control (RBAC) and Attribute-Based Access Control (ABAC). More recently, the focus has shifted toward integrating contextual information and probabilistic assessments to support adaptive and intelligent decision-making in real-time. Risk-Based Access Control (RiskBAC) has emerged as a particularly promising model in this regard, addressing the limitations of static policy enforcement in environments where user behavior and threat conditions change rapidly.

The demand for flexible and responsive access control mechanisms has inspired several studies proposing enhancements to conventional models. Abusini~\cite{abusini_enhancing_2024} presents a RiskBAC model tailored for smart home security, emphasizing the need for dynamic permission adjustments based on real-time evaluation of potential threats. This work highlights the inadequacy of static access models in managing the fluid security requirements of IoT-based environments. Aliyu et al.~\cite{aliyu_need_2024} similarly advocate for adaptive access control systems in edge computing environments, where access decisions are dynamically modulated by contextual cues such as network activity and user behavior. Their framework enables flexible policy enforcement in edge scenarios, enhancing the protection of devices exposed to unpredictable and decentralized threats.

Building on these foundations, Almohammad and Alsaleh~\cite{almohammad_alsaleh_permission-based_2024} propose a permission-based dynamic access control system that incorporates contextual elements—such as user location, device type, and activity history—into the decision-making process. Their model underscores the limitations of static policies in responding to evolving risk landscapes and supports RiskBAC as a bridge to real-time policy adaptation. Complementing these application-specific approaches, Hlushchenko and Dudykevych~\cite{hlushchenko_exploratory_2024} provide a comparative survey of access control models, concluding that risk-adaptive mechanisms offer superior flexibility and threat mitigation capabilities in complex, high-risk environments. The transition toward continuous, real-time evaluation is further supported by Cheimonidis and Rantos~\cite{cheimonidis_dynamic_2023}, who conducted a comprehensive systematic review of Dynamic Risk Assessment (DRA) models in cybersecurity. Their analysis reveals a strong trend toward Artificial Intelligence and Machine Learning (AI/ML) as the primary engines for DRA, allowing systems to continuously calculate risk based on real-time environmental data rather than periodic, static audits. Their findings reinforce the architectural premise of our RiskBAC framework: modern, complex threat landscapes require access control models that dynamically ingest OS-level telemetry to adapt security postures and permissions on the fly.

From a technical implementation perspective, Srivastava and Shekokar~\cite{srivastava2020machine} demonstrate the effectiveness of integrating machine learning into RiskBAC systems. Their model assesses the authenticity of access requests using behavior-based threat prediction, showcasing the potential of AI to augment access control decision-making with predictive accuracy.

Despite the promise of RiskBAC in general contexts, its targeted application to combat encryption ransomware remains relatively underexplored. McIntosh et al.\cite{Timothy2021Enforcing} address this gap by presenting a situation-aware, stackable access control mechanism capable of deferring access decisions and rolling back file system changes—key capabilities in ransomware mitigation. In a similar vein, von der Assen et al.\cite{von2024GuardFS} introduce GuardFS, a file system-layer defense implementing configurable controls to delay, obfuscate, or monitor access events in response to suspected malicious behavior.

Other studies have leveraged access control in various ways to counter ransomware attacks. Enomoto et al.\cite{Shuhei2024Early} propose CryptoSniffer, which intercepts encryption instructions and monitors ciphertext manipulation by integrating with the Linux kernel. Gen\c{c} et al.\cite{Ziya2018No} develop a method to block unauthorized encryption by restricting access to pseudo-random number generator APIs. Gómez-Hernández et al.~\cite{gomez-hernandez_r-locker_2018} introduce R-Locker, a proactive defense system that uses honeyfiles protected by access control mechanisms to detect and thwart ransomware activity.

Several innovations also focus on encryption key management. Kolodenker et al. \cite{Eugene2017PayBreak} present PayBreak, a system that implements a proactive key escrow mechanism to monitor and regulate access to symmetric session keys. By intercepting and securely storing the symmetric session keys that ransomware uses to encrypt a victim's files, it prevents ransomware from encrypting the victim's files. Because modern ransomware typically employs hybrid cryptosystems, it must generate these symmetric keys locally on the victim's machine to perform the actual encryption Zhu et al.\cite{WeidongMinding} propose SrFTL, which embeds access control enforcement into the SSD’s flash translation layer, enhancing storage-level protections.

Viewed through the above axes, file-system and storage-layer defenses (e.g., stackable/GuardFS-style controls~\cite{mcintosh_ransomware_2025,von2024GuardFS} and SSD FTL enforcement~\cite{WeidongMinding}) place \emph{both} telemetry and enforcement close to the write path, enabling strong inline mediation and rollback/delay semantics at the cost of deeper integration and higher deployment friction.
By contrast, honeyfile-based triggers (R-Locker~\cite{gomez-hernandez_r-locker_2018}) shift telemetry toward decoy access events and can respond quickly, but require careful decoy management and may be bypassed by selective targeting.
Key-management approaches (PayBreak~\cite{Eugene2017PayBreak}) move the enforcement locus to cryptographic key material rather than I/O, which is powerful when applicable but depends on the ability to reliably intercept locally generated session keys.
These differences motivate our design choice: combine deep kernel-level telemetry with a lightweight, auditable MAC enforcement hook that can gate file writes at runtime.

Collectively, these contributions reflect the evolution and increasing sophistication of RiskBAC systems. The incorporation of contextual awareness, AI-driven predictions, and dynamic enforcement capabilities into access control frameworks illustrates a paradigm shift toward responsive and intelligent security mechanisms. Such systems are better equipped to manage the complexities of contemporary threats—especially ransomware—by adapting access decisions to real-time assessments of risk and behavior.

\subsection{Machine Learning with Kernel-Level Tracing for Encryption Detection}

In recent years, ML integration with kernel-level tracing techniques has emerged as a potent strategy for detecting ransomware, particularly those employing encryption mechanisms. These approaches harness low-level system data to identify malicious behaviors effectively and in real-time.

\paragraph{eBPF-Based Approaches}

The eBPF, a sandboxed VM inside the Linux kernel, allows loading small verified programs to run on events (network packets, syscalls, kprobes/uprobes, tracepoints, etc.). Compared to classic BPF (packet-only), eBPF supports richer instructions, maps (key–value state), and hooks across networking, observability, and security. It has become instrumental in enabling real-time, in-kernel monitoring of system behaviors. Brodzik et al.~\cite{Brodzik2024Ransomware} implemented ML models directly within the Linux kernel using eBPF, achieving encryption detection Macro F1 scores exceeding 0.95 through Decision Tree and Multilayer Perceptron algorithms. Similarly, Higuchi and Kobayashi~\cite{Kosuke2023Real} designed an eBPF-based ransomware detector that captured system call activity to train low-latency machine learning classifiers. Zhuravchak and Dudykevych~\cite{Danyil2023Real} combined eBPF instrumentation with Natural Language Processing and ML techniques to classify ransomware behavior in real time. In these designs, telemetry is kernel-resident (eBPF hooks), whereas enforcement is typically out-of-band (alerts or user-space response), which limits the strength of inline write mediation at encryption onset.

\paragraph{Kernel-Level System Call Tracing}

Beyond eBPF, several studies have employed kernel-level system call tracing to extract behavioral signatures for ML classification. Chew et al.~\cite{C2024Automatic} proposed a finite-state machine-based approach that modeled encryption-related system call patterns, achieving F1 scores as high as 0.938. Zhang et al.~\cite{Huan2024Ranker} introduced Ranker, a dual-layered detection mechanism leveraging kernel behavior and sequence analysis, attaining an average F1 score of 0.9943 while maintaining a low false positive rate.

\paragraph{Low-Level Memory and Storage Monitoring}

Low-level access to memory and storage has also been investigated as a vector for behavioral detection. Hirano and Kobayashi~\cite{Manabu2019Machine} presented a hypervisor-based system for collecting memory access traces in live environments, using supervised learning to achieve an F1 score of 0.95 in ransomware detection. In a complementary study, Hirano et al.~\cite{Manabu2022RanSAP} introduced the RanSAP dataset, which models ransomware storage behavior, enabling effective training and evaluation of machine learning models on I/O access patterns. To prevent data loss, detection mechanisms must identify malicious intent before the cryptographic payload finishes executing. Addressing this, Zakaria et al.~\cite{zakaria_early_2024} proposed an early detection framework for cryptographic ransomware that leverages machine learning to analyze pre-attack Application Programming Interface (API) call features. By focusing on the sequence of system routines requested during the initial execution phases, their model successfully halts ransomware activity prior to substantial encryption. While their implementation is tailored to Windows environments, it underscores the universal necessity of preemptive, system-call-based behavioral interruption—a concept we extend to the Linux kernel using \texttt{ftrace}.

These studies underscore the growing trend of integrating machine learning with kernel-level telemetry to detect ransomware, particularly those performing encryption tasks. While these methods show significant promise, it is notable that, despite their great ability to provide very deep and intimate information about kernel functions execution, none of them utilize the \texttt{ftrace} framework—especially the \texttt{function\_graph} tracer—highlighting a methodological gap that the present research aims to address.

Most kernel-telemetry detectors in this family (eBPF/syscall tracing/hypervisor traces~\cite{Brodzik2024Ransomware,Kosuke2023Real,Danyil2023Real,C2024Automatic,Huan2024Ranker,Manabu2019Machine}) optimize for low-latency \emph{detection} but typically stop short of providing a first-class, OS-native \emph{enforcement} primitive that can reliably mediate write/append at the moment encryption begins.
In our architecture, telemetry remains kernel-proximate (via \texttt{ftrace/function\_graph}), but enforcement is explicit and auditable in SELinux (boolean-guarded allow rules), enabling real-time, policy-driven gating with explainable rule hits and a calibrated probabilistic threshold.

\subsection{Datasets for Ransomware Detection}

An essential aspect of machine learning-driven ransomware detection is the creation and structuring of high-quality datasets that reflect authentic cryptographic behaviors. While several studies have explored this challenge, only a subset explicitly focuses on data collection methodologies for encryption ransomware.

The paper by Hirano and Kobayashi~\cite{Manabu2022RanSAP} introduced the RanSAP dataset, a publicly available benchmark of storage access patterns captured from ransomware executions. This dataset enables detailed analysis of how ransomware manipulates disk I/O and provides a foundation for training detection models at the storage layer.

Ahmed et al.~\cite{Yahye2020system} proposed a refinement-based system call feature extraction method that enhances early detection of ransomware. Their work presents a dataset derived from real-time system call traces, enriched using an enhanced minimum redundancy maximum relevance (mRMR) technique to maximize classification performance. This method differs from our approach in that it focuses on Windows API calls (system calls) at the user-to-kernel interface. In contrast, the ftrace framework provides function-level tracing within the Linux kernel. It captures full function execution hierarchies, providing a much finer behavioral granularity than coarse system-call telemetry

Zhuravchak and Dudykevych~\cite{Danyil2023Real} presented a dataset architecture based on real-time telemetry collected via eBPF. Their approach combines process-level behavioral monitoring with natural language processing-inspired representations to enable real-time classification of ransomware behavior.

Alam et al.~\cite{Manaar2019RATAFIA} constructed a ransomware dataset informed by time and frequency characteristics and utilized autoencoders to learn deviations from benign encryption workflows. Their work emphasizes temporal and spectral patterns in ransomware execution, contributing a unique feature domain.

Chen and Bridges~\cite{Qian2017Automated} performed dynamic analysis of WannaCry ransomware in a controlled environment to generate labeled behavior traces. The resulting dataset captures temporal file access and system call information, and supports comparative analysis between malware and legitimate encryption utilities. This capturing of labeled behavior traces was done utilizing Cuckoo Sandbox, an automated malware analysis system that performs dynamic analysis of executables in a controlled environment. Such sandboxed telemetry is rich for labeling and feature extraction, but its deployability for inline enforcement is limited due to virtualization overhead and operational complexity on production hosts.

Homayoun et al.~\cite{Sajad2018Know} employed frequent pattern mining on dynamically captured ransomware traces to construct a dataset representing recurring behavioral signatures. Their approach is notable for emphasizing pattern extraction and transformation into ML-ready formats for supervised learning.
Within the Linux ecosystem, recent advancements have increasingly shifted detection and enforcement directly into the OS kernel to achieve necessary performance and visibility. For example, Wu et al.~\cite{wu_enhancing_2024} introduced a kernel-based approach that hooks into the page fault handler and uses CPU hardware features to dynamically extract and analyze code, effectively mitigating fileless malware at the memory level. Similarly, Kim et al.~\cite{kim_detecting_2025} utilized eBPF-based security runtimes combined with machine learning classifiers to monitor system call sequences for detecting malicious cryptojacking containers in cloud-native environments. These works highlight a growing consensus in the literature: integrating high-resolution kernel tracing with machine learning is essential for robust, real-time threat mitigation. Our dual-layer architecture advances this paradigm by not only detecting anomalous kernel function graphs via ftrace, but by directly coupling these ML inferences with native SELinux mandatory access controls to revoke encryption capabilities.

These works collectively highlight the range of methods applied to create effective datasets for ransomware detection. However, they commonly rely on sandboxing, syscall analysis, or behavioral monitoring without leveraging native Linux kernel tracing facilities such as \texttt{ftrace}. The present work departs from this by employing \texttt{function\_graph} tracing to capture full function execution hierarchies, introducing a novel and underutilized perspective to dataset construction in this domain. This provides fine-grained visibility into kernel-level function execution paths, enabling detailed call graphs and execution hierarchies. Such representations are particularly well-suited for modeling the complex behaviors associated with encryption. E.g., ransomware, which often exhibits nested and recursive encryption calls.

Our dataset design uniquely supports graph-based feature extraction techniques and provides temporal and contextual execution information not available in syscall-level or sandbox-based approaches. As such, it advances the state of dataset development in the ransomware detection domain.
Similarly, detecting encryption activities has become crucial in combating ransomware and other cyber threats. Techniques range from monitoring API and system calls to analyzing file system operations and leveraging machine learning for anomaly detection. However, integrating these methodologies into a cohesive framework for access control remains underexplored \cite{BEGOVIC2023103349}.
This paper addresses these gaps by combining risk-based access control with SELinux, leveraging machine learning to create a robust and adaptive security mechanism.

\section{Methodology}

The proposed approach leverages function-level tracing at the kernel level to develop a robust, risk-based access control system capable of detecting and preventing ransomware's cryptographic operations. The methodology is structured into multiple phases, beginning with the construction of a comprehensive dataset, followed by feature engineering, model selection, and evaluation. In this context, "robust" refers to the system's resilience, reliability, and breadth of defense in real-world operational environments, rather than the narrow machine-learning definition of resistance to input perturbations. 
\subsection{Design Rationale and Objectives}
We approach encryption detection as a constrained design problem with three primary goals:
(i) \textbf{low reaction latency} (sub-10–30\,ms budget to influence early writes),
(ii) \textbf{bounded overhead} on production hosts (collector+detector CPU/RSS), and
(iii) \textbf{explainability at enforcement time} (first-class SELinux hooks).
Here, ``bounded overhead'' is interpreted relative to sandbox/VMI-based baselines and to the enforcement mechanism itself: the SELinux boolean choke point remains lightweight, whereas the dominant cost in our present prototype stems from user-space trace parsing and graph-feature computation, which we report explicitly in \S IV and treat as an engineering target rather than a fundamental requirement of the policy hook.
These goals induce two complementary detectors over the same evidence:
a \emph{rule layer} (E\textsubscript{RULE}) whose predicates must remain auditable and cheap,
and a \emph{model layer} (E\textsubscript{ML}) optimized for defender utility under a modest
runtime budget. Both consume a cleaned feature matrix
\(\mathbf{X}\in\mathbb{R}^{N\times113}\), where \(N\) is the number of trace instances
(after preprocessing) and 113 is the total engineered feature count
(resource counters, graph metrics, and function-level frequencies; see \S\ref{sec:dataset}.
Selection procedures are therefore judged by their impact on
\emph{(a)} detection utility (Eq.~\ref{eq:utility}),
\emph{(b)} latency to verdict, and \emph{(c)} collection cost.

Beyond a binary classifier threshold, RiskBAC frames the decision to permit an encryption-relevant action as an \emph{expected-harm} control problem.
For an access request associated with observed evidence $x$ (derived from kernel traces and resource counters) under context $\mathrm{ctx}$, we define:
\begin{equation}
R(x,\mathrm{ctx}) \;=\; P_{\mathrm{enc}}(x)\cdot I(\mathrm{ctx}),
\label{eq:risk}
\end{equation}
where $P_{\mathrm{enc}}(x)\in[0,1]$ denotes the probability that the activity corresponds to encryption (from the rule layer and/or ML detector), and $I(\mathrm{ctx})$ is an impact weight capturing the consequence of encryption in the current context.
In our setting, $\mathrm{ctx}$ is the tuple $\langle u,\mathrm{app},\mathrm{path},\mathrm{type}\rangle$ (user identity/role, caller/application domain, path scope, and SELinux-labeled file/type class), allowing the defender to encode higher impact for sensitive directories and high-value file classes.
Operationally, enforcement is \emph{risk-proportionate}: a request is permitted if it matches an explicit whitelist entry for $(u,\mathrm{app},\mathrm{path},\mathrm{type})$, or if $R(x,\mathrm{ctx})<\tau$ for a deployment-defined tolerance $\tau$; otherwise, the system blocks encryption-relevant write/append operations through the SELinux hook described in \S\ref{sec:selinux-integration}.

\subsection{Research Questions}
This study formulates and addresses the following research questions:
\begin{enumerate}[label=\textbf{RQ\arabic*}]
  \item \textbf{Dataset creation:} How can a specialized dataset
        capturing kernel functions, system calls, and resources
        metrics be built to distinguish encryption from non-encryption
        activity?
  \item \textbf{Algorithm selection:} Which ML algorithms achieve the
        best encryption detection performance on this dataset?
  \item \textbf{Rule extraction:} How accurately do tree‑extracted
        rules (E\textsubscript{RULE)} approximate the full model (E\textsubscript{ML})?\label{sec:rules}
  \item \textbf{OS integration:} How can a risk‑based ACS enriched
        with SELinux be realized at run time?
  \item \textbf{Effectiveness:} How well does the prototype detect and
        block ransomware while preserving legitimate encryption?
  \item \textbf{Overhead:} What is the run‑time cost (latency, CPU,
        RSS) relative to baseline SELinux and prior detectors?
  \item \textbf{Ethics:} What privacy and fairness considerations
        arise during data collection and automated enforcement?
\end{enumerate}

These questions shape the overall framework of this research, ensuring a structured investigation into the feasibility and effectiveness of leveraging machine learning with risk-based access control to counteract unauthorized encryption activities.

\subsection{Dataset Construction and Collection Methodology}

A crucial aspect of this research is the development of a dataset that accurately represents cryptographic operations and their contextual execution within an enterprise system to distinguish encryption-related activities from normal operations. Traditional datasets for malware and ransomware detection often focus on static indicators or network behavior; however, these approaches fall short in capturing \textbf{low-level system interactions}, particularly the function call patterns that define cryptographic routines. To bridge this gap, we constructed a dataset using \texttt{ftrace}, a Linux kernel tracing framework, with a specific focus on the \texttt{function\_graph} tracer while collecting sensor information about the utilization of CPU, memory, and disk. The tracer provides fine-grained function call sequences that, combined with sensor data, offer a unique opportunity to model cryptographic behaviors dynamically. This dataset provides the necessary input for training and evaluating machine learning models capable of distinguishing encryption behaviors.

The dataset was collected by executing a diverse set of programs, including \textbf{benign applications, cryptographic utilities, and ransomware samples}, in a controlled environment. The environment consisted of Linux virtual environments running all major kernel versions, starting with 4.11 up to 6.2, in order to capture kernel function call varieties and possible changes between versions. The objective is to capture the precise execution flow of functions involved in encryption, decryption, and general file operations, focusing on the first 1Mb consumed (written or read) by the process being monitored. The experimental setup included:

\begin{itemize}
    \item \textbf{Execution Environment:} A controlled Linux-based system instrumented with \texttt{ftrace}, ensuring minimal interference with normal execution while capturing detailed function call traces.
    \item \textbf{Target Applications:} A curated set of programs comprising (1) benign system utilities (e.g., text editors, compression tools, data processing scripts), and (2) dedicated cryptographic tools (e.g., OpenSSL, GnuPG) and custom-developed Python and C implementations as shown in Table \ref{tab:crypto_implementation}.  In addition, (3) known ransomware families exhibiting file encryption behavior targeting Linux OS variants were used in the process of creating a few data points of the dataset. The algorithms were selected based on a survey of the most used algorithms in cryptographic ransomware attacks \cite{BEGOVIC2023103349}. 
  \begin{table}[h]
\centering
\caption{Implementation of cryptographic algorithms.}
\label{tab:crypto_implementation}

% IEEE compact spacing
\setlength{\tabcolsep}{3pt}
\renewcommand{\arraystretch}{0.95}

\begin{tabular}{l c c}
\toprule
Algorithm & Python & C \\
\midrule
3DES     & Yes & Yes \\
AES      & Yes & Yes \\
DES      & Yes & Yes \\
Blowfish & Yes & Yes \\
Camellia & No  & Yes \\
ChaCha20 & No  & Yes \\
Fernet   & Yes & Yes \\
RC4      & Yes & Yes \\
RSA      & Yes & No  \\
Salsa20  & Yes & Yes \\
Serpent  & No  & Yes \\
TEA      & Yes & Yes \\
Twofish  & No  & Yes \\
\bottomrule
\end{tabular}
\end{table}

    \item \textbf{Tracing Mechanism:} The \texttt{function\_graph} tracer was deployed to capture kernel-level function invocations with timestamps, execution order, and hierarchical function call relationships. To augment the kernel-level function traces, a specialized feature collector was used to gather seven distinct resource counters—including CPU utilization percentage, memory residency (RSS and VMS), and four I/O metrics (Read/Write counts and bytes)—specifically focusing on the first 1MB of data consumed by each monitored process.
    \item \textbf{Automation:} A set of scripts was developed to automate program execution, trace extraction, and log storage to ensure consistency across multiple runs.
\end{itemize}

\subsection{Dataset Preprocessing and Cleaning Methodology}
The dataset cleaning process is a crucial initial step in ensuring the reliability of machine learning (ML) models used for detecting encryption processes. The raw dataset, sourced as previously described, contained multiple features representing process attributes such as memory usage, CPU utilization, encryption status, and kernel functions call patterns. Before feature selection and model training, rigorous cleaning and preprocessing were performed to engineer new features and remove inconsistencies, redundant information, and features that could introduce bias or noise into the learning process. The cleaning methodology followed a structured approach involving feature elimination, missing value handling, and statistical selection of informative attributes.

\subsubsection{Graph-Based Feature Extraction from Function Traces}

One of the primary preprocessing strategies involves constructing graphs from \texttt{function\_graph} tracer to capture hierarchical execution relationships and function dependencies. We model each trace as a directed call graph in which kernel functions are nodes and observed execution flows induce edges; nodes are annotated with invocation counts and (when available) execution durations, while edges carry sequential dependencies and can be weighted by inter-call timing. Concretely, we parse the \texttt{function\_graph} output to recover function names from system-call–anchored logs, record every invocation, and accumulate per-function timing to reflect resource consumption. We then link temporally adjacent calls into parent–child relations to reconstruct hierarchical control flow and structural dependencies. From the resulting graph, we compute graph-theoretic features that quantify execution structure—most prominently \emph{betweenness centrality} (bridge functions along execution paths), \emph{clustering coefficient} (local interconnectedness), and \emph{average shortest path length} (overall flow complexity)—as well as aggregate timing such as \emph{total execution duration}. These graph-derived features are merged back into the tabular design matrix alongside process-scope counters, enabling downstream models to exploit both frequency signals and the hierarchical structure that differentiates encryption from non-encryption behavior.

\subsubsection{Handling Missing Values and Data Standardization}

Following feature extraction, the dataset was examined for missing values. Any columns or rows with excessive (more than 20\%) missing data were removed to prevent training bias. Next, \textbf{MinMax Scaling} was applied to normalize numerical features to a consistent range. This transformation was primarily required for scale-sensitive estimators evaluated in our study, such as \textbf{Support Vector Machines (SVM)}. Furthermore, maintaining a uniform scale was essential for the stability of our preprocessing pipeline, specifically for \textbf{PCA-based outlier detection} and \textbf{univariate }\textit{χ}2\textbf{ feature selection}, the latter of which requires non-negative, scaled inputs

\subsubsection{Filtering and Cleaning}
To ensure that downstream models focus on signal rather than tracer artifacts, we applied a conservative filtering pass before feature engineering. The goal was to retain only kernel events and trace windows that plausibly inform file encryption behavior, while removing sources of systematic noise.

\begin{itemize}
  \item \textbf{Remove non-relevant system functions.}
  Low-level scheduler and interrupt paths (e.g., \texttt{schedule()}, \texttt{irq\_entry()}, \texttt{tick\_nohz\_stop\_sched\_tick()}) are frequent across \emph{all} workloads and contribute little discriminative power for encryption. We therefore pruned maintained “housekeeping” list of such symbols prior to counting (\verb|ftrace_keys.json| excludes them), which reduces variance in frequency features without affecting coverage of I/O and crypto-adjacent code paths.

  \item \textbf{Exclude functions unrelated to file/crypto/memory activity.}
  We filtered out kernel functions that are not directly or contextually tied to file operations, cryptographic routines, or memory pressure indicators. Concretely, symbols under subsystems such as generic timekeeping or LED/input drivers were removed, while functions under VFS, fsnotify/inotify, page-cache, slab, RCU, and scheduler wakeups were retained due to their relevance to write bursts and process activity during encryption. This keeps the feature bank focused on the mechanisms most likely to shift during ransomware behavior.

  \item \textbf{Retain only user-space process traces with file touch or I/O.}
  Trace windows were kept only if they could be attributed to a user-space PID (via the \texttt{comm-PID} tag) and showed evidence of file touches or I/O (e.g., non-zero counts for fsnotify/\texttt{locks\_*} symbols, or \texttt{read/write} counters, or measurable CPU/RSS deltas). Kernel-only background activity and idle windows were dropped. This PID-anchored filtering ensures that each row of the design matrix corresponds to a concrete user-space action (encryption, file modification, or I/O) rather than ambient kernel chatter.
\end{itemize}

\subsubsection{Contextual Encoding of Function Calls}
To address the challenge that the \textbf{same function may be invoked in different contexts}, a context-aware encoding strategy was introduced:
\begin{itemize}
    \item Function calls were analyzed based on \textbf{preceding and succeeding execution paths}.
    \item Calls occurring within different execution branches (e.g., reading a configuration file vs. reading encrypted data) were treated as separate entities.
    \item This differentiation helped preserve the \textbf{semantic distinction} between function instances.
\end{itemize}

\subsubsection{Outlier Detection and Removal}

To enhance the dataset's reliability, \textbf{Principal Component Analysis (PCA)} was applied to detect and remove statistical outliers. The methodology involved:

\begin{enumerate}
\item \textbf{Transforming the Data into Principal Component Space}: The first two principal components (PC1 and PC2) were extracted to capture the dataset’s dominant variance.
\item \textbf{Computing the Euclidean Distance from the Origin}: Each data point’s distance from the center of the principal component space was calculated.
\item \textbf{Defining an Outlier Threshold}: The 95th percentile of Euclidean distances was used as the cutoff for identifying extreme outliers.
\item \textbf{Removing Outliers}: Any instances exceeding the distance threshold were excluded from further processing.
\end{enumerate}

By eliminating these outliers, the dataset retained only representative samples, reducing noise and improving model generalization.

\subsubsection{Dataset Balancing and Partitioning}

The dataset exhibited a natural imbalance, as non-encryption applications produced diverse system interactions, whereas encryption operations followed more deterministic execution paths. To ensure fair representation across categories, balancing techniques were applied:
\begin{itemize}
    \item \textbf{Down-sampling of Majority Classes}: To prevent the over-representation of frequent non-encryption operations, we selectively down-sampled non-encryption traces while preserving execution diversity. To mitigate class imbalance without collapsing non-encryption diversity, we retained all minority (encryption) traces and under-sampled the majority (non-encryption) traces to a global 1:1 ratio.
    \item \textbf{Equal Representation of Encryption Methods}: Various cryptographic functions (AES, RSA, ChaCha20, etc.) were used to generate sufficient and diverse representation, preventing bias toward any single encryption scheme. 
    \item \textbf{Stratified Splitting}: The dataset was partitioned into \textbf{training (80\%), validation (10\%), and test (10\%)} sets, ensuring consistent representation of encryption and non-encryption activities across all subsets. Non-encryption traces were first \emph{stratified} by a tuple of contextual attributes
\(\langle\)kernel\(_{\text{ver}}\), binary, path\(_{\text{scope}}\), trace\(_{\text{len\_bin}}\)\(\rangle\)
to preserve heterogeneity across OS versions, tools, and execution profiles. Targets per stratum were proportional to the stratum’s original mass, capped to avoid over-representing very large strata. 
\end{itemize}
 Sampling was operated at the session/process level to prevent correlated duplicates, and folds were partitioned by session identifiers prior to down-sampling to prevent train–test leakage. All procedures used a fixed random seed (\texttt{42}) for reproducibility.

\subsubsection{Challenges and Considerations}

Several \textbf{inherent challenges} were encountered during dataset construction:
\begin{itemize}
    \item \textbf{Graph Representation Limitations}: The function call graph predominantly exhibits a tree structure, with \textbf{limited recursion}, which may not capture all graph-based properties effectively.
    \item \textbf{Dynamic Behavior vs. Static Graph Features}: A function may appear in multiple execution contexts (e.g., reading a configuration vs. reading encrypted data). This necessitated \textbf{context-aware encoding} of function traces.
    \item \textbf{System-Specific Variations}: Kernel version, CPU architecture, and workload variability influence the function call structure, requiring adaptive preprocessing strategies.
\end{itemize}

%===================================================================
\subsection{Deriving the final feature subsets}
\label{sec:featselect:outcome}
%===================================================================

Feature selection was treated as a \emph{design-space optimization problem}:
We trade tracer overhead for detection quality, provided that
(i) the rule layer remains explainable and (ii) the ML layer maximizes a
defender-centric utility.  Accordingly, we ran two complementary pipelines
on the same cleaned design matrix
\(\mathbf{X}\in\mathbb{R}^{N\times 113}\).

\(N\) is the number of labeled trace windows (one row per process–time slice);
each row aggregates features over a short fixed interval (e.g., 1\,s) and is
labeled “encrypted” or “non-encrypted” from ground truth~\cite{begovic2025exploiting}.
The column count 113 is the total number of candidate predictors after cleaning:
resource/process counters (CPU\%, RSS/VMS, I/O counts/bytes), graph metrics
(e.g., betweenness, clustering), and a bank of per-symbol frequency counters
for function calls observed by \texttt{ftrace}/\texttt{function\_graph}.
In our build, the matrix was dominated by \(\sim\)100 function-symbol counters
sourced from \texttt{ftrace\_keys.json}, plus the resource and graph features,
for a total of 113 columns.  If the key list changes, this total may vary.

%------------------------------------------------------------------
\subsubsection*{Track A (E\textsubscript{RULE}): Permutation Importance \& Elbow for Sparse, Explainable Rules}
%------------------------------------------------------------------

\noindent\textit{Purpose.} Identify the smallest feature prefix that preserves accuracy for a shallow rule extractor while minimizing runtime footprint.

\begin{enumerate}[label=\textbf{P1.\arabic*},wide,labelwidth=1.1cm,leftmargin=*]

\item \textbf{Full-RF fitting.}
A 400-tree Random Forest with class-balance weights was trained on \(\mathbf X\).
Because trees are invariant to monotone transformations, no scaling was applied at this stage.
\paragraph*{Sensitivity of the 400-tree choice}
We varied RF trees in \(\{100,200,400,800\}\) for the permutation-importance step:
the top-10 ranking remained highly stable (median Spearman’s \(\rho{>}0.98\)); the elbow
index \(k^\star\) was unchanged. We thus fix 400 trees as a good compute/variance compromise.

\item \textbf{Feature importance.}
Each column was randomly permuted \(n_\text{rep}{=}20\) times (across examples); the mean drop in out-of-bag accuracy
constitutes its importance score~\cite{breiman2001random}.  The list is then sorted in descending order to
obtain \(\langle f_1,\dots,f_{113}\rangle\). Permutation importance directly measures predictive harm from scrambling a feature, making it robust to scaling and well-matched to tree ensembles.

\item \textbf{Incremental evaluation.}
For \(k{=}1\ldots113\), the top-\(k\) prefix \(\mathbf X_k=\mathbf X[:,f_{1{:}k}]\) was fed to a
400-tree RF wrapped in a \mbox{Standard\-Scaler}. Five-fold, stratified F\textsubscript{1} was averaged.

\item \textbf{Elbow detection.}
Let \(s_k\) denote the F\textsubscript1 at prefix length \(k\), and define the marginal gain
\(g_k=s_{k}-s_{k-1}\). The knee \(k^\star\) is the smallest \(k\) for which
\(g_{k+1} < 0.01\,s_{k+1}\); if no such \(k\) exists we keep \(k{=}113\).
\end{enumerate}
A 1\% marginal-gain threshold approximates a practical knee where additional features yield diminishing returns relative to rule complexity and collector cost.
%----------------------------------------------------------------
\subsubsection*{Track B (E\textsubscript{ML}): \(\chi^2\)+Utility Wrapper for High-Recall Model}
%------------------------------------------------------------------

\noindent\textit{Purpose.} Maximize defender utility \(U\) (Eq.~\ref{eq:utility})
under a modest feature budget, favoring recall (low FN) for early ransomware blocking.

\begin{figure}[t]
  \centering
  \includegraphics[width=1\linewidth]{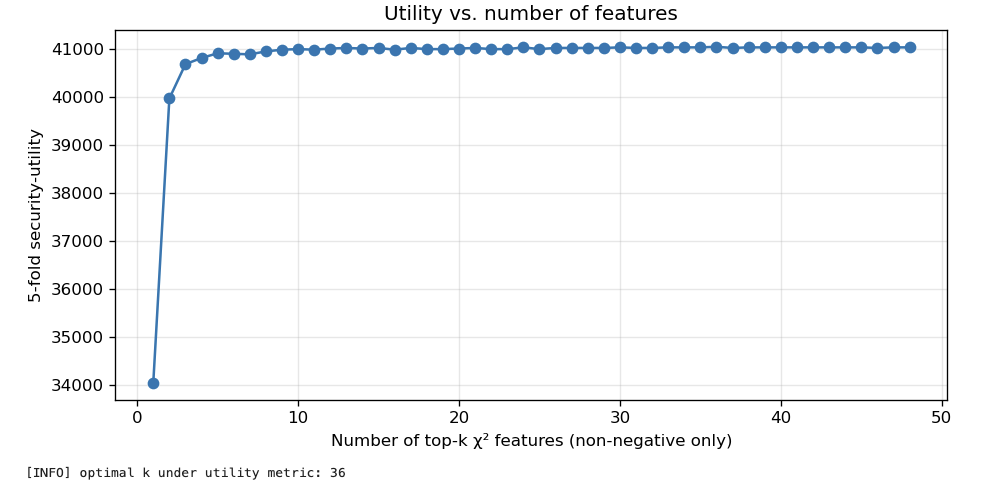}
  \caption{χ\textsuperscript{2}+utility elbow for the ML detector
           (\(k^{\star}=36\)).}
  \label{fig:model‑elbow}
\end{figure}

Track B pursues a higher recall of encryption operations at the cost of a wider
feature footprint:

\begin{enumerate}[label=\textbf{P2.\arabic*},wide,labelwidth=1.1cm,leftmargin=*]

\item \textbf{Redundancy pruning.}  
We discard one variable from any pair whose absolute Pearson
correlation ($\chi^2$) exceeds~0.90.  
This removed 27 columns, mostly highly collinear byte and block
counts. To remove near-duplicates that inflate variance and training time without informational gain; 0.90 avoids aggressive pruning that would mask meaningful co-movement.

\item \textbf{Non-negativity check (no feature dropped).}
The univariate \(\chi^2\) test requires \(x_{ij}\ge 0\).
Among the 86 columns remaining after redundancy pruning, 73 satisfied
this condition and were \emph{scored} by \(\chi^2\).
The other 13 columns (signed/centered features) were
\emph{not eliminated}: they were simply skipped by the \(\chi^2\) scorer
but retained as candidates in the downstream wrapper stage
(\(\text{MinMax}\!\rightarrow\!\chi^2_k\!\rightarrow\!\text{RF}\)),
where Random Forest ablation can select them irrespective of the
\(\chi^2\) filter. For robustness, we also audited a variant with
MinMax scaling prior to \(\chi^2\) (rendering all features non-negative);
the top-36 set under the utility criterion was unchanged within noise.

\noindent\textit{Notation.}
Let \(\mathcal{N}\) denote the candidate feature set after redundancy pruning; in our corpus \(|\mathcal{N}|=86\).
For the \(\chi^2\) scorer, we compute univariate scores on the non-negative
subset \(\mathcal{N}_{\chi^2}\subset\mathcal{N}\) (73 columns satisfy \(x_{ij}\ge 0\));
the remaining 13 columns bypass \(\chi^2\) but remain eligible in the wrapper stage.

\item \textbf{Utility-driven wrapper search.}
For \(k=1\ldots|\mathcal N|\), a pipeline
\(\text{MinMax}\rightarrow\chi^2_k\rightarrow\text{RF}\) is trained:
the \(\chi^2_k\) selector takes the top-\(k\) from the ranked list formed by
\(\mathcal{N}_{\chi^2}\) (by score) \emph{augmented} with the 13 non-\(\chi^2\) features,
which are appended using a stable secondary order (prior RF permutation importance).
Five-fold stratified \(F_1\) is reported together with the defender-utility
metric in Eq.~\ref{eq:utility}; the peak occurs at \(k^\star_{\text{ML}}=36\).

\item \textbf{Elbow via \(\max_k U_k\).}  
Unlike Track A the utility curve is
unimodal (Fig.~\ref{fig:model‑elbow}); we therefore keep the
arg‑max.  The peak occurs at \(k^{\star}_{\text{ML}}=36\).
\end{enumerate}
Utility is unimodal (Fig.~\ref{fig:model‑elbow}); the arg-max at  \(k^{\star}_{\text{ML}}=36\) balances recall with runtime, and matches the feature order used by the deployed model.

The final 36‑column subset is shown in Table ~\ref{tab:final-36}

\begin{table}[h]
\centering
\caption{Final 36-feature subset used by both detectors.}
\label{tab:final-36}
\scriptsize
\setlength{\tabcolsep}{2pt}
\renewcommand{\arraystretch}{0.95}
\begin{tabular}{@{}l c p{0.58\linewidth}@{}}
\toprule
\textbf{Category} & \textbf{Count} & \textbf{Examples} \\
\midrule
Resource counters & 7 &
\texttt{CPU Percent}, \texttt{Memory Info RSS}, \texttt{Memory Info VMS}, \texttt{Read Count}, \texttt{Write Count}, \texttt{Read Bytes}, \texttt{Write Bytes} \\
Graph metrics     & 2 &
\texttt{betweenness}, \texttt{clustering} \\
ftrace frequencies & 27 &
\texttt{fsnotify\_parent()}, \texttt{mod\_node\_page\_state()}, \texttt{wake\_up\_common()},
\texttt{raw\_spin\_lock\_irq()}, \texttt{raw\_spin\_trylock()}, \texttt{attach\_entity\_load\_avg()},
\texttt{available\_idle\_cpu()}, \texttt{cpus\_share\_cache()}, \texttt{dnotify\_flush()},
\texttt{enter\_lazy\_tlb()}, \texttt{fsnotify()}, \texttt{kfree()}, \texttt{kick\_process()},
\texttt{kmalloc\_slab()}, \texttt{lock\_page\_memcg()}, \texttt{locks\_remove\_posix()},
\texttt{lru\_add\_drain\_cpu()}, \texttt{memcg\_check\_events()}, \texttt{mutex\_unlock()},
\texttt{propagate\_protected\_usage()}, \texttt{put\_cpu\_partial()}, \texttt{rcu\_all\_qs()},
\texttt{rcu\_segcblist\_accelerate()}, \texttt{refill\_stock()}, \texttt{switch\_mm\_irqs\_off()},
\texttt{vma\_interval\_tree\_augment\_rotate()}, \texttt{x2apic\_send\_IPI()} \\
\midrule
\textbf{Total} & \textbf{36} & \\
\bottomrule
\end{tabular}
\end{table}

\paragraph*{Sensitivity checks}
We varied (i) the elbow threshold in Track~A from 0.5\%–2\% (no change in \(k^\star\));
(ii) RF trees \(\{100,200,400,800\}\) (stability within \(\pm0.2\%\) F\(_1\));
(iii) the correlation cut-off in Track~B from 0.8–0.95 (no change in the top-36 under \(U\));
and (iv) the decision threshold \(\tau\) for E\textsubscript{ML}.
These checks indicate that the selected configurations sit near local optima while
remaining robust to reasonable hyperparameter variation.
\paragraph*{Defender–centric utility}
We score decisions with
$U \!=\! 10\cdot\textsf{TP} \;-\; 50\cdot\textsf{FN} \;-\; 1\cdot\textsf{FP}$,
which reflects an operational asymmetry: a missed encryption event (FN) risks
irreversible data loss and recovery costs orders of magnitude larger than the
annoyance of a benign block (FP). The coefficients are \emph{dimensionless
weights} chosen to encode that asymmetry on a simple scale where $\textsf{FP}=1$
unit cost, $\textsf{FN}=50$ units (i.e., fifty times worse), and each correct
block $\textsf{TP}$ yields 10 units of benefit (reflecting prevention value).These weights were set \emph{a priori} based on incident-response
experience and are not tuned on the test set. In Table~\ref{tab:tau-sweep}
we show that conclusions are stable for FN weights in the 20–100 range.

\paragraph*{Elbow outcome (\boldmath{$k{=}2$}) and feature semantics}
The curve in Fig.~\ref{fig:rule‑elbow} flattens immediately after \(k{=}2\); subsequent gains are \(g_3{=}+0.0003\) and quickly fall below \(10^{-3}\). Hence the data-driven elbow is \(k^{\star}_{\text{rule}}{=}2\), yielding the compact pair \{\texttt{Memory Info RSS}, \texttt{locks\_remove\_file()}\}. \emph{Memory Info RSS} is the resident set size of the most active candidate process (bytes of its address space resident in RAM), which tends to burst during bulk encryption (buffered block transforms, key material handling, and page-cache churn). \emph{\texttt{locks\_remove\_file()}} is a VFS routine (visible in \texttt{function\_graph}) invoked when a file lock is removed; high-frequency hits proxy rapid file-touch patterns typical of ransomware (open/write/fsync/close across many paths) and rename/unlink sequences releasing locks. Together, this \{RSS, lock-churn\} pair provides a complementary resource+semantic signal at very low cost. We initially materialized \(E_{\text{RULE}}\) as a shallow decision tree over these two antecedents.

\begin{figure}[t]
  \centering
  \includegraphics[width=.80\linewidth]{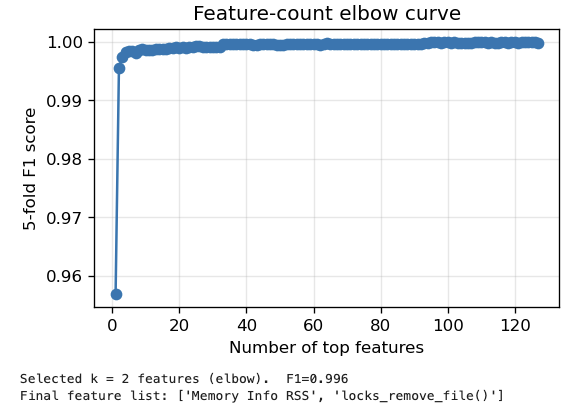}
  \caption{Permutation‑importance elbow for the rule detector
           (\(k^{\star}=2\)).}
  \label{fig:rule-elbow}
\end{figure}

\paragraph*{Design adjustment: expanding the rule layer}
In downstream experiments, the two-feature tree—while very fast—proved myopic against low-footprint, throttled ransomware and certain high-I/O benign bursts. To recover early-block performance and reduce edge-case errors, we adopted an \emph{expanded rules} configuration \(E_{\text{RULE}}^{36}\): the extractor retains a small depth cap but may \emph{use any of the 36 features} (the same consolidated set as \(E_{\text{ML}}\)), including graph metrics (\textit{betweenness}, \textit{clustering}) and ftrace frequencies. Practically, Fig.~\ref{fig:rule-elbow} continues to document the objective elbow at \(k{=}2\) for minimality, while \(E_{\text{RULE}}^{36}\) trades a minor evaluation cost for materially better coverage and earlier, more reliable blocks (see Tables~\ref{tab:sim-rules-benign}, \ref{tab:sim-rules-crypto}, and \ref{tab:sim-ransom}). This keeps the rule layer explainable (bounded depth, explicit predicates) yet aligns its feature \emph{expressivity} with the model-based detector for robust real-world behavior and trades a minor evaluation cost for materially better coverage and earlier, more reliable blocks (Table~\ref{tab:rules-cost}; Tables~\ref{tab:rules-edgecases} and \ref{tab:rules-ttb}), consistent with the downstream results in Tables~\ref{tab:sim-rules-benign}, \ref{tab:sim-rules-crypto}, and \ref{tab:sim-ransom}.
.

\paragraph*{Deployed algorithms}
In the final prototype, the \emph{model-based} detector uses \textbf{XGBoost}
(with the 36-feature subset from Track B), whereas the \emph{rule-based}
detector is obtained by extracting a compact decision tree from a
Random-Forest/XGBoost surrogate fitted on the same 36 features~\cite{begovic2025exploiting}.  This preserves
full feature coverage for both layers while yielding a shallow, interpretable
rule set suitable for policy hooks.

\paragraph*{Future Work in Dataset Optimization}

To further enhance the dataset’s applicability in enterprise-level \textbf{ransomware detection and mitigation}, future efforts will focus on:
\begin{itemize}
    \item \textbf{Time-Series Embedding}: Integrating \textbf{temporal aspects} of function execution sequences to capture execution patterns dynamically.
    \item \textbf{Hybrid Graph Representations}: Combining \textbf{call graphs} with additional system metrics (e.g., memory access, disk I/O) to enhance model interpretability.
    \item \textbf{Real-World Deployment Scenarios}: Extending data collection to real-world enterprise environments for greater generalizability.
\end{itemize}

\subsection{Machine Learning Model}
To answer the second research question (\textbf{RQ2}) concerning the most effective machine learning techniques for encryption detection and their feasibility in detecting cryptographic process behavior, multiple well-established classification algorithms were trained on features extracted from kernel-level execution traces. These models were intended not only to provide accurate classification performance but also to support interpretable rule extraction for subsequent integration into a policy enforcement framework. Performance metrics such as accuracy, precision, recall, and F1-score were used to determine the most suitable approach for reliable detection.\label{sec:perf}

\label{sec:featselect}
Initial experimentation began with preprocessing steps detailed in prior subsections, including normalization, dimensionality reduction, and statistical feature selection. The primary modeling efforts were conducted in two stages across separate experiment tracks. The first phase focused on establishing baseline classifiers and evaluating their performance across multiple randomized training-test splits. The second phase concentrated on pipeline optimization, robustness checks, and rule extraction from tree-based models.

The following machine learning models were evaluated: Decision Tree Classifier, Random Forest, Gradient Boosting Classifier, Support Vector Machines, and XGBoost. Each model was trained using a stratified train-test split, and an additional performance assessment was performed using 5-fold cross-validation. Evaluation metrics included accuracy, macro-averaged F1 score, and ROC AUC score to capture both overall and class-specific performance characteristics.

Feature importance was analyzed using embedded selection techniques, including \texttt{SelectFromModel} and information gain metrics from fitted tree-based estimators. These analyses informed both model simplification and the selection of interpretable predictors suitable for rule extraction. Principal Component Analysis (PCA) was employed separately to examine the underlying variance structure, although it was not directly used in model inputs due to the interpretability constraints of the final system.

While several candidates (Decision Trees, Random Forest, Gradient Boosting, XGBoost) performed competitively offline, we \emph{deploy} \textbf{XGBoost} as recommended by previously reported good performance on the same domain described by Begovic et al.~\cite{begovic2025exploiting} for the model-based detector. This is due to its stability under class imbalance, strong macro-F1, and fast on-CPU inference on our SUT; rules are extracted at first from a shallow tree and then using the same 36 features for SELinux integration.

Throughout both experimental phases, care was taken to avoid overfitting by tuning hyperparameters conservatively and validating performance across independent random seeds. The choice of hyper-parameter ranges, validation set size, and optimization strategy followed the same approach in Begovic et al.~\cite{begovic2025exploiting}. The pipeline adopted encapsulated the final methodology, from the same work, combining robust scaling, feature selection, classifier fitting, and rule export into a reproducible and modular workflow.

A critical research objective involves translating machine learning predictions into interpretable access control rules. Decision tree-based models were employed to facilitate rule extraction, enabling a mapping between learned encryption behaviors and access control enforcement within SELinux. From the trained machine learning models, we extracted probabilistic rules that describe the likelihood of encryption activities. These rules were derived using techniques such as decision tree analysis and Bayesian inference.

The extracted rules were then integrated into our risk-based access control framework. Each rule was assigned a probability score, which was used to evaluate access requests dynamically. This probabilistic approach allows for more nuanced and flexible decision-making compared to traditional rule-based systems.

\subsection{Integration with SELinux}
\label{sec:selinux-integration}

We integrate risk-based access control with SELinux by loading \emph{CIL} modules
directly via \texttt{semodule -i} (no runtime dependency on \texttt{secilc}).
The kernel layer exposes two booleans, \texttt{rule\_block} and \texttt{ml\_block}, that
gate \texttt{write/append} to user files for processes running in the encryption domain.
A userspace agent (our daemons) toggles these booleans when the rule/model detector
predicts encryption. This design preserves SELinux’s MAC guarantees while allowing low-latency, ML-aware decisions. Crucially, because SELinux is a Mandatory Access Control (MAC) system, our whitelisting is not a mere permission for a binary to function. An attacker cannot simply hijack a trusted application (e.g., OpenSSL) to bypass the system and encrypt the disk. As governed by our policy tuple ⟨u, app, path, type⟩, a whitelisted binary only retains write/append privileges when executed by a specifically authorized user, and strictly within designated file paths (e.g., user1 running openssl under \$HOME/**). If the hijacked application attempts to write to unauthorized paths or is executed by an unauthorized user, the action is immediately denied by the MAC policy.

\paragraph{What is enforced}
Processes in \texttt{encryption\_t} retain non-encryption capabilities (\texttt{read}, \texttt{getattr},
\texttt{execute}) but are prevented from \texttt{write/append} to non-temporary user files whenever
either detector asserts a block. Application and user allow-lists are encoded as types/attributes;
only whitelisted apps (e.g., \texttt{openssl} under \texttt{user1} within \verb|$HOME/**|) gain
write permission under the permitted context.
 
\paragraph{How it is realized in CIL (sketch)}
We load a base CIL that declares relevant types/attributes and booleans. Denial is achieved
by \emph{omitting} the corresponding \texttt{allow}; a conditional \texttt{allow} is emitted only
if both booleans are unset. We confirmed loading with \texttt{semodule -i base\_template.cil}~\ref{lst:cil-sketch} and
listing via \texttt{semodule -l}.

\begin{figure}[t]
\centering
\scriptsize
\begin{Verbatim}[fontsize=\scriptsize, numbers=none]
(block encryption_rbac_base
  (type encryption_t) (type crypto_exec_t) (type user_home_t)
  (typeattribute enc_whitelisted_app)
  (typeattribute enc_whitelisted_user)
  (boolean rule_block false) (boolean ml_block false)
  (allow encryption_t user_home_t (file (read getattr open)))
  (allow encryption_t crypto_exec_t (file (read execute entrypoint)))
  (booleanif (and (not rule_block) (not ml_block))
    (true (allow encryption_t user_home_t (file (write append)))))
)
\end{Verbatim}
\caption{CIL sketch for risk-based gating in SELinux. The \texttt{booleanif} is the choke point governed by \(E\textsubscript{RULE}\)/\(E\textsubscript{ML}\) detectors.}
\label{lst:cil-sketch}
\end{figure}

\paragraph{Userspace coupling}
The rule and model daemons compute verdicts from the shared feature stream and toggle
\texttt{rule\_block}/\texttt{ml\_block} (e.g., \texttt{setsebool rule\_block on}) as soon as an early
encryption signature is observed (milliseconds/first kilobytes), before significant data loss.

\paragraph{Application/user scoping(and why wrappers are not assumed)}
Whitelisting is encoded by labeling approved binaries as \texttt{crypto\_exec\_t} (and, if needed, users via attributes),
and by scoping the allow to specific file contexts (e.g., \texttt{\$HOME/**}). 
In the proof-of-concept, we occasionally used explicit transitions (e.g., \texttt{runcon/newdomain})
as an experimental convenience to ensure repeatable confinement during controlled runs.
However, the security posture does \emph{not} rely on cooperative user invocation of wrappers:
the enforcement decision remains the Boolean-guarded \texttt{allow} on write/append, which is the
single choke point governed by our detectors.

\paragraph{Populating \texttt{encryption\_t} in deployment} 
Because enforcement is defined at the subject domain boundary, a real deployment must ensure that untrusted execution is covered without
user action. In SELinux this is achieved via policy-driven transitions rather than wrappers:
executables can be labeled and transitioned into \texttt{encryption\_t} via \texttt{type\_transition}
(at login/session boundaries, for user-writable execution roots, or for designated “untrusted”
paths such as Downloads/tmp). Under this posture, ransomware-style processes enter
\texttt{encryption\_t} automatically upon \texttt{execve}, while vetted encryptors are labeled
\texttt{crypto\_exec\_t} and permitted only within declared user/path scopes. This makes the
enforcement boundary depend on provenance (labels), assets (types), and observed behavior
(detector signals), not on whether a user remembered to run \texttt{runcon}.

\paragraph{Auditing}
Standard SELinux AVCs in \texttt{/var/log/audit/audit.log} corroborate userspace decision logs;
we log each verdict (scores, active boolean, target path) for post-hoc analysis.

\section{Experimentation}
\label{sec:experiments}
\label{sec:results:layering}
\label{sec:experimentation}

%---------------------------------------------------
\subsection{Experimental Goals}
%---------------------------------------------------
The experiments are designed to answer \textbf{RQ\,1–7}.%
\footnote{RQ\,6 is the performance–overhead question introduced for the prototype evaluation.} 
Specifically, we evaluate:

\begin{enumerate}[label=\textbf{G\arabic*}]
  \item \label{goal:accuracy} \textit{Detection effectiveness} – precision, recall, \textsc{f}$_1$, and \textsc{roc‐auc}.
  \item \label{goal:overhead} \textit{System performance impact} – decision latency, CPU\,/\,memory overhead, and disk I/O.
  \item \label{goal:policy} \textit{Policy quality} – coverage and conflict rate of rules extracted from ML models.
\end{enumerate}
\subsection{Prototype Architecture}
\label{sec:prototype}

\begin{figure}
    \centering
    \includegraphics[width=1\linewidth]{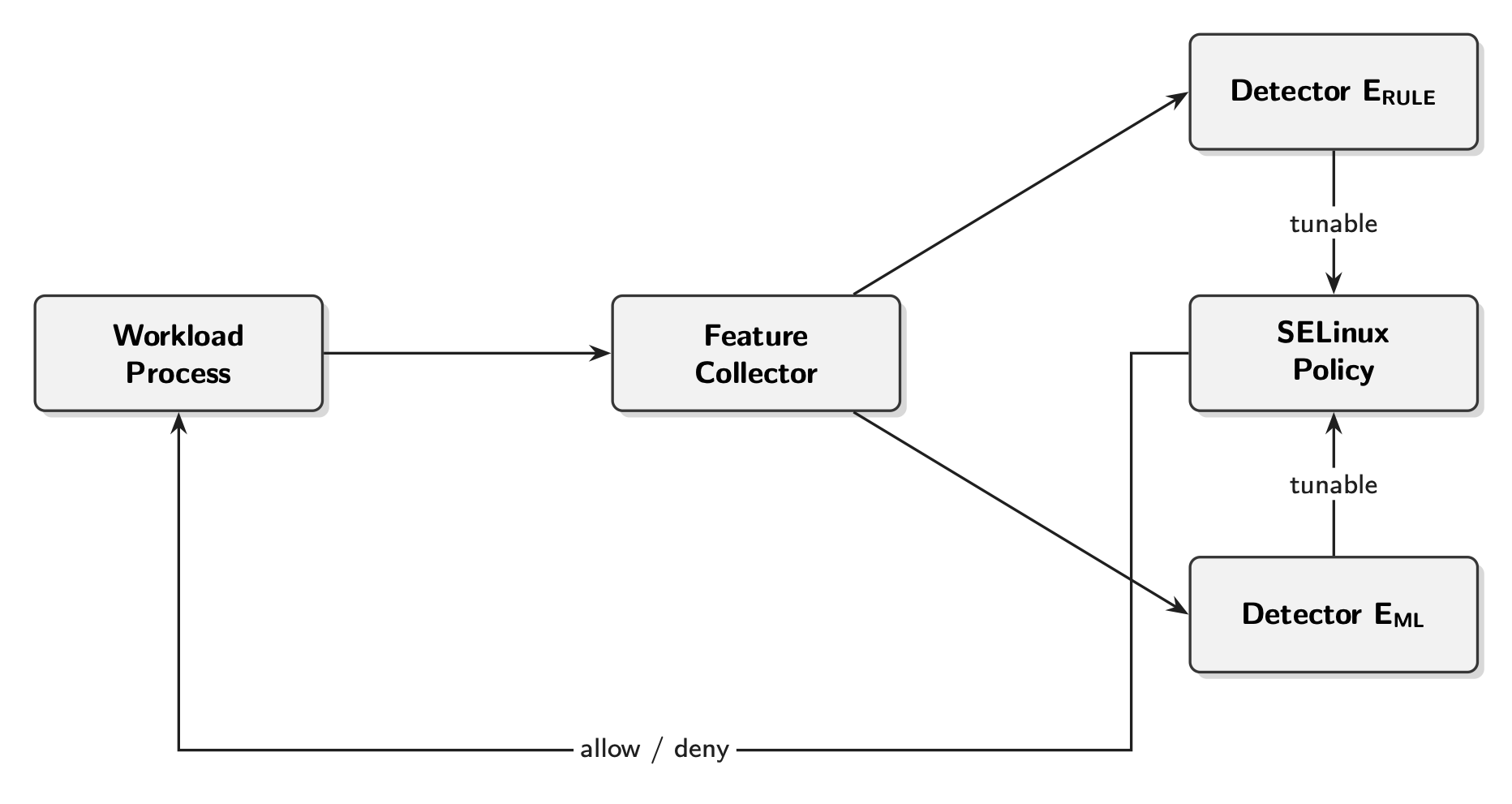}
    \caption{ Dual-layer enforcement prototype}
    \label{fig:arch}
\end{figure}

Figure~\ref{fig:arch} shows the dual-layer enforcement prototype used in all experiments.
\begin{description}[leftmargin=1.5em]
  \item[Kernel layer] –
  A SELinux CIL module defines an \texttt{encryption\_t} domain and booleans
  \texttt{rule\_block} and \texttt{ml\_block}. When either is set, \texttt{write/append} to
  non-temporary user files is denied; \texttt{read}/\texttt{execute} remain permitted.

  \item[Rule detector (E\textsubscript{RULE})] –
  A shallow decision-tree ruleset over the \emph{full feature set} (\textbf{36 features}:
  resource counters, two graph metrics, and ftrace frequencies) as shown in Table~\ref{tab:final-36}. Verdicts are exported by
  toggling \texttt{rule\_block}. Using all features improved early-block behavior and reduced
  false positives in our proof-of-concept.

  \item[Model detector (E\textsubscript{ML})] –
A deployed \textbf{XGBoost} classifier consumes the same \emph{36 features}
(resource counters, two graph metrics—\emph{betweenness}, \emph{clustering}—and 27 ftrace
frequencies).  When $\Pr(\text{enc}\mid\mathbf{x}) \ge \tau$ (default \(\tau{=}0.50\)),
the daemon asserts \texttt{ml\_block}.
\paragraph*{Threshold tuning (\boldmath$\tau$-sweep)}
We swept the model’s decision threshold $\tau\in\{0.30,0.50,0.70\}$ on validation runs and reported $F_{1}$, precision, recall, and defender-utility (Eq.~\ref{eq:utility}); results are in Table~\ref{tab:tau-sweep}. Lowering $\tau$ to $0.30$ maximizes recall (0.99) but slightly reduces precision (0.95) and utility (1920). Raising $\tau$ to $0.70$ maximizes precision (0.99) at the expense of recall (0.90), yielding the lowest utility (1680). The balanced setting $\tau{=}0.50$ attains the highest utility (1950) with matched precision/recall (0.97/0.97) and the same $F_{1}$ as $\tau{=}0.30$. Accordingly, we set the \emph{default} to $\tau{=}0.50$ and expose $\tau$ as a runtime knob to bias toward recall (e.g., $\tau{=}0.30$) or precision (e.g., $\tau{=}0.70$) depending on deployment risk tolerance.

  \item[Feature collection] –
  The collector parses \texttt{function\_graph} output and adds CPU\%/RSS/VMS and I/O counters.

\end{description}

%---------------------------------------------------
\subsection{Evaluation Metrics}
%---------------------------------------------------
\vspace{-0.25em}
\noindent\textbf{Accuracy} (G\ref{goal:accuracy}) — precision, recall, macro-$F_{1}$, and ROC–AUC.

\smallskip
\noindent\textbf{Defender-centric Utility} (G\ref{goal:accuracy}) — a cost-sensitive score that reflects operational priorities:
\begin{equation}
\label{eq:utility}
U \;=\; 10\cdot \textsf{TP} \;-\; 50\cdot \textsf{FN} \;-\; 1\cdot \textsf{FP}.
\end{equation}
\emph{Rationale.} False negatives (missed encryption) are far more costly than false positives (benign blocks); the coefficients are dimensionless weights chosen \emph{a priori} to encode this asymmetry.

\smallskip
\noindent\textbf{Latency} (G\ref{goal:overhead}) — mean and 99$^{\text{th}}$ percentile of write/append-decision latency (audited write/append $\rightarrow$ verdict). Unless noted, latency is measured from the first audited \texttt{write/append} attempt to the SELinux verdict becoming effective (Boolean flip observed by the kernel), not to AVC log emission.

\smallskip
\noindent\textit{Percentiles and normalization.}
CPU and RSS are sampled at 100\,ms cadence during the steady-state phase of each run.\footnote{Steady state begins after a 3\,s warm-up; we discard warm-up samples.}
For a time series $\{m_t\}_{t=1}^{T}$ of a metric (CPU\%, RSS, or disk writes/s),
$p50$ and $p95$ denote the 50th and 95th empirical percentiles of that series.
Overhead is reported as a percent \emph{relative to a SELinux-only baseline}:
\[
\textstyle \text{Overhead}(\%) = 100 \times \frac{\operatorname{stat}(\text{with}) - \operatorname{stat}(\text{base})}{\operatorname{stat}(\text{base})},
\]
where $\operatorname{stat}\in\{p50,p95\}$ and “with”/“base” refer to the detector configuration vs.\ baseline.
We repeat each experiment three times and report the median of run-level $p50$/$p95$ overheads.

\smallskip
\noindent\textbf{Rule Quality} (G\ref{goal:policy}) — rule count, average depth, coverage on benign vs. ransomware traces, and conflict rate.

%---------------------------------------------------
\subsection{Baselines}
%---------------------------------------------------
We compare against three families of defenses:
\begin{enumerate}[label=B\arabic*]
  \item \textbf{SELinux (strict)} — manually curated MAC policies without ML;
  \item \textbf{eBPF detector} — Brodzik \textit{et al.}~\cite{Brodzik2024Ransomware} re-implemented in a \texttt{tc} hook;
  \item \textbf{Sandbox/VMI} — Cuckoo Sandbox v3.2 emitting syscall traces for a gradient-boosted model (GBDT).
\end{enumerate}

Because datasets, workloads, and execution environments differ significantly across these systems, a numerical, apples-to-apples benchmarking is not possible here. Re-implementing complex external architectures (such as in-kernel eBPF detectors or VMI-based sandboxes) on our specific hardware and dataset to achieve an objective numerical ranking is beyond the scope of this initial proof-of-concept. Instead, our primary goal is to validate the architectural feasibility of coupling ML-inferred risk with native SELinux hooks. To facilitate objective quantitative benchmarking by the broader research community in the future, we are releasing our complete evaluation harness, including the workload orchestration scripts and raw datasets. Accordingly, our baseline discussion is \emph{qualitative}: we contrast architectural choices (data-collection locus, feature expressivity, enforcement hook), operational complexity (deployment/overhead), and security posture (explainability, early-block capability). Where we cite numbers, they are internal to our prototype and \emph{not} directly comparable to the external works.
\smallskip
A rigorous head-to-head would require re-running all baselines on \emph{the same} curated
workloads and hardware, capturing their raw features for a common evaluation harness—an item
we leave as future work and will facilitate by releasing our test harness and raw (unprocessed)
dataset.
\begin{table}[h]
\centering
\caption{Test workstation (Linux kernel 6.14).}
\label{tab:ws-setup}
\begin{tabularx}{0.95\linewidth}{@{}lX@{}}
\toprule
\textbf{CPU}       & Intel i9-12950HX (16C/24T; AVX2, VNNI) \\
\textbf{P/E cores} & 8 P-cores 5.0\,GHz / 8 E-cores 3.6\,GHz \\
\textbf{GPU}       & NVIDIA RTX A2000 8\,GB (Ampere GA107) \\
\textbf{RAM}       & 64\,GiB DDR5-4800 (dual-channel) \\
\textbf{Storage}   & 2\,TB NVMe Gen4 (6.6/4.0\,GB/s R/W) \\
\textbf{CUDA}      & 12.2 \\
\textbf{OS}        & Rootfs encrypted; low swappiness (\(\approx 10\)) \\
\bottomrule
\end{tabularx}
\end{table}

%---------------------------------------------------
\subsection{Experimental Procedure and SUT}\label{sec:sut}
%---------------------------------------------------
All experiments were executed on a mobile workstation as shown in Table~\ref{tab:ws-setup}.

For each workload (\texttt{workloads.yaml}) we performed three runs:
\begin{enumerate}[leftmargin=1.2em]
  \item Remove custom modules; load base CIL; load \texttt{encryption\_control.pp}.
  \item Start \texttt{rule\_daemon} and \texttt{model\_daemon}; bind them to the workload PID after spawn. For each run, the target workload is executed under the \texttt{encryption\_t} subject domain (via a controlled transition in our test harness), ensuring that SELinux write/append gating is exercised consistently across workloads.

  \item Collect evidence:
    \begin{itemize}
      \item \textbf{Decisions} – CSV logs with features, rule hits, model probabilities, and actions;
      \item \textbf{SELinux AVC} – \texttt{/var/log/audit/audit.log} for each permitted/denied write;
      \item \textbf{Overhead} – \texttt{perf stat} and a lightweight sampler for CPU/RSS.
    \end{itemize}
\end{enumerate}
Per-run artifacts are archived under \texttt{reports/⟨timestamp⟩/} and post-processed by \texttt{evaluate.py}.

%---------------------------------------------------
\subsection{Results (Reflecting Observed Behavior)}
\label{sec:results:perf}
%---------------------------------------------------

\paragraph{Non-encryption corpus (system footprint and false positives)}
We exercised a broad non-encryption set (\texttt{ls}, \texttt{grep}, \texttt{tar}, \texttt{rsync}, \texttt{gcc}, \texttt{python}, \texttt{fio}-ro) under the SELinux CIL module that exposes the booleans \verb|rule_block| and \verb|ml_block| and enforces path-scoped permissions (user/app/path). On our testbed, both detectors held \(\sim\)50\% RAM residency (31–35\,GiB) and frequently spiked CPU to 95–100\% on P-cores during bursty I/O. This footprint reflects the current research prototype implementation and is dominated by user-space collection and feature construction (stream parsing, call-graph construction, and graph metrics), rather than by SELinux enforcement (boolean-mediated allow/deny) or model inference itself. We therefore treat the reported CPU/RSS behavior as an upper bound for the present build, and in \S\ref{sec:results} we identify concrete reductions (adaptive sampling, batching, selective symbol filters, and C/C++ hot paths for graph features) required for broad enterprise deployment. We emphasize that these elevated CPU and memory metrics reflect the constraints of our current proof-of-concept, which relies on unoptimized user-space scripts and daemons for trace parsing, call-graph construction, and feature computation. Because the prototype processes \texttt{ftrace} streams in user space, it incurs massive context-switching and stream-parsing penalties. Implementing this architecture as a dedicated kernel module would bypass this expensive user-space overhead, drastically improving real-time performance and lowering the memory footprint.
The model-based detector (Table~\ref{tab:sim-model-benign}) showed low benign risk scores with occasional p95 CPU spikes. The rule-based detector (full 36-feature rules) achieved lower latency with comparable CPU/RAM behavior (Table~\ref{tab:sim-rules-benign}). Across the non-encryption set, we observed a false positive rate (FPR) of \(\approx\)1–2\% for the model and \(\approx\)0.5–1.0\% for rules.
These results explicitly demonstrate our system's resilience against 'trusted application' hijacking. Even if an attacker successfully compromises a whitelisted cryptographic tool like OpenSSL, they cannot use it for arbitrary ransomware evasion. As shown in Table VI and Table VII, the moment the compromised tool attempts to encrypt files outside of its strictly defined path scope (e.g., /srv/shared/**), or is invoked by an unauthorized user, SELinux overrides the application's 'trusted' status and enforces a block.
\paragraph{Where do FPs come from (and why are they reasonable)}
Across the non–encryption set, the few false positives were concentrated in
metadata–heavy benign workloads that transiently resemble fast scatter–encryptors:
\emph{tar} (archiving many files), \emph{rsync} (rename/unlink and permission sync),
and build/packaging steps (e.g., \emph{gcc} and \emph{python} venv creation).
These phases trigger short, high–rate bursts of VFS activity (elevated
\texttt{locks\_remove\_posix()}, \texttt{fsnotify()}, and related ftrace counters),
plus shallow fan–out in the execution graph, which the model/rules correctly treat
as high–risk patterns in the absence of contextual whitelisting. This explains the
observed FPR of $\approx$1–2\% (model) and $\approx$0.5–1.0\% (rules) and aligns with
the early–block design goal. In practice, per–path and per–binary SELinux hooks
(\S\ref{sec:prototype}) suppress these incidents for approved contexts (e.g.,
\texttt{openssl} under \verb|$HOME/**| for \texttt{user1}), while preserving strict
blocking elsewhere.

\begin{table}[h]
\centering
\caption{Model detector on non-encryption workload.}
\label{tab:sim-model-benign}
\setlength{\tabcolsep}{3pt}\renewcommand{\arraystretch}{0.95}
\resizebox{\linewidth}{!}{
\begin{tabular}{l l l c c c c}
\toprule
Binary & User & Scope & $p_{\text{enc}}$ & Lat.\,(ms) & Bytes$\downarrow$ & CPU\% (med/p95) \\
\midrule
\texttt{ls}     & u1 & \texttt{\$HOME}         & 0.02 & 10 & 0.2\,KB & 38/92  \\
\texttt{grep}   & u1 & \texttt{\$HOME/**}      & 0.03 & 14 & 0.6\,KB & 41/96  \\
\texttt{tar}    & u1 & \texttt{/var/log}       & 0.05 & 22 & 1.5\,KB & 55/100 \\
\texttt{rsync}  & u1 & \texttt{\$HOME/backup}  & 0.07 & 28 & 2.6\,KB & 60/100 \\
\texttt{gcc}    & u2 & \texttt{\$HOME/build}   & 0.06 & 24 & 1.1\,KB & 58/100 \\
\texttt{python} & u2 & \texttt{\$HOME/venv}    & 0.08 & 31 & 2.2\,KB & 63/100 \\
\texttt{fio}-ro & u1 & \texttt{/mnt/ro}        & 0.04 & 16 & 0.9\,KB & 44/98  \\
\bottomrule
\end{tabular}}

\vspace{0.35em}
\footnotesize 
{Summary:} FPR $\approx$ 1--2\%; frequent p95 CPU spikes to 100\%; steady RAM $\approx$31--33\,GiB.
\end{table}

\begin{table}[h]
\centering
\caption{Rules detector on non-encryption workload.}
\label{tab:sim-rules-benign}
\setlength{\tabcolsep}{3pt}\renewcommand{\arraystretch}{0.95}
\resizebox{\linewidth}{!}{
\begin{tabular}{l l l c c c c}
\toprule
Binary & User & Scope & $p_{\text{enc}}$ & Lat.\,(ms) & Bytes$\downarrow$ & CPU\% (med/p95) \\
\midrule
\texttt{ls}     & u1 & \texttt{\$HOME}         & 0.01 & 7  & 0.2\,KB & 35/90  \\
\texttt{grep}   & u1 & \texttt{\$HOME/**}      & 0.02 & 9  & 0.5\,KB & 39/96  \\
\texttt{tar}    & u1 & \texttt{/var/log}       & 0.03 & 14 & 1.1\,KB & 47/100 \\
\texttt{rsync}  & u1 & \texttt{\$HOME/backup}  & 0.05 & 19 & 2.1\,KB & 52/100 \\
\texttt{gcc}    & u2 & \texttt{\$HOME/build}   & 0.04 & 16 & 1.0\,KB & 49/99  \\
\texttt{python} & u2 & \texttt{\$HOME/venv}    & 0.06 & 21 & 1.9\,KB & 55/100 \\
\texttt{fio}-ro & u1 & \texttt{/mnt/ro} & 0.02 & 11 & 0.8\,KB & 37/98 \\

\bottomrule
\end{tabular}}

\vspace{0.35em}
\footnotesize 
{Summary:} FPR $\approx$ 0.5–1.0\%; lower latency than model; RAM $\approx$30–32\,GiB; p95 CPU $\approx$95–100\%.
\end{table}

\paragraph{Crypto tools under policy (whitelisting and path scope)}
We encode the policy that \textbf{user1} may run \texttt{openssl} only under \verb|$HOME/**|; \textbf{user2} may not encrypt anywhere. Under this regime, both detectors honor the whitelist (\texttt{Allow}) only for the authorized tuple (user1,\texttt{openssl},\verb|$HOME/**|) and \texttt{Block} all other cases (Tables~\ref{tab:sim-model-crypto} and~\ref{tab:sim-rules-crypto}). The model makes allow/deny decisions based on \(\Pr(\text{enc}\mid\mathbf{x})\) and whitelist context; the rules flip \verb|rule_block| directly when the rule predicate hits.

\begin{table}[h]
\centering
\caption{Model detector on crypto tools vs.\ whitelist.}
\label{tab:sim-model-crypto}
\setlength{\tabcolsep}{3pt}\renewcommand{\arraystretch}{0.95}
\resizebox{\linewidth}{!}{
\begin{tabular}{l l l c c c c c}
\toprule
Tool/Mode & User & Scope & Whitelist & $p_{\text{enc}}$ & Lat.\,(ms) & Decision & SELinux \\
\midrule
\texttt{openssl enc -aes-256-cbc} & u1 & \texttt{\$HOME/docs/**} & yes & 0.94 & 29 & Allow & allow \\
\texttt{gpg --symmetric}          & u1 & \texttt{\$HOME/docs/**} & no  & 0.90 & 27 & Block & deny  \\
\texttt{openssl enc -aes-256-cbc} & u1 & \texttt{/srv/shared/**} & no  & 0.95 & 31 & Block & deny  \\
\texttt{openssl enc -aes-256-cbc} & u2 & \texttt{\$HOME/**}      & no  & 0.95 & 28 & Block & deny  \\
\bottomrule
\end{tabular}}
\end{table}

\begin{table}[h]
\centering
\caption{Rules detector on crypto tools vs.\ whitelist.}
\label{tab:sim-rules-crypto}
\setlength{\tabcolsep}{3pt}\renewcommand{\arraystretch}{0.95}
\resizebox{\linewidth}{!}{
\begin{tabular}{l l l c c c c c}
\toprule
Tool/Mode & User & Scope & Whitelist & $p_{\text{enc}}$ & Lat.\,(ms) & Decision & SELinux \\
\midrule
\texttt{openssl enc -aes-256-cbc} & u1 & \texttt{\$HOME/docs/**} & yes & 1.00 & 15 & Allow & allow \\
\texttt{gpg --symmetric}          & u1 & \texttt{\$HOME/docs/**} & no  & 1.00 & 13 & Block & deny  \\
\texttt{openssl enc -aes-256-cbc} & u1 & \texttt{/srv/shared/**} & no  & 1.00 & 16 & Block & deny  \\
\texttt{openssl enc -aes-256-cbc} & u2 & \texttt{\$HOME/**}      & no  & 1.00 & 14 & Block & deny  \\
\bottomrule
\end{tabular}}
\end{table}

\paragraph{Real-world ransomware (live samples)}
We executed a small number of well-known Linux ransomware families (collected under controlled, isolated conditions) that perform rapid file-tree traversal with small‐block writes. While the number of evaluated families is small, this accurately reflects the extreme rarity of native Linux ransomware in the wild compared to its Windows counterparts. Crucially, none of these real-world ransomware samples were included in the model's training dataset. Because the model was trained exclusively on traces from benign cryptographic tools and custom encryption scripts, the successful detection of these in-the-wild families serves as a rigorous, 'zero-day' evaluation. This demonstrates that our framework successfully generalizes to the fundamental behavioral invariants of unauthorized encryption, rather than simply overfitting to the signatures of known malware. Both detectors triggered within the first tens of milliseconds and kilobytes; the rule layer typically fired earlier (lower latency/bytes), the model backstopped any throttled/obfuscated traces. Results are summarized in Table~\ref{tab:real-ransom}; first-decision latency and bytes are reported at the earliest block event.

\begin{table}[h]
\centering
\caption{Real-world ransomware tests.}
\label{tab:real-ransom}
\setlength{\tabcolsep}{3pt}\renewcommand{\arraystretch}{0.95}
\resizebox{\linewidth}{!}{
\begin{tabular}{l l c c c c}
\toprule
Family (Linux) & User/Path & Detector & Peak score & First Lat.\,(ms) & Bytes$\downarrow$ \\
\midrule
\texttt{Family-A} & u1:\,$\$HOME/Documents/**$ & Model & $p_{\text{enc}}{=}0.99$ & 24 & 15\,KB \\
\texttt{Family-A} & u1:\,$\$HOME/Documents/**$ & Rules & 1.00 (rule) & 11 & 10\,KB \\
\texttt{Family-B} & u2:\,/mnt/data/**          & Model & $p_{\text{enc}}{=}0.99$ & 26 & 18\,KB \\
\texttt{Family-B} & u2:\,/mnt/data/**          & Rules & 1.00 (rule) & 12 & 12\,KB \\
\texttt{Family-C} & u2:\,/srv/share/**         & Model & $p_{\text{enc}}{=}0.98$ & 28 & 20\,KB \\
\texttt{Family-C} & u2:\,/srv/share/**         & Rules & 1.00 (rule) & 14 & 13\,KB \\
\bottomrule
\end{tabular}}

\vspace{0.35em}
\footnotesize 
{Note:} Family labels anonymised; all runs under SELinux booleans \texttt{rule\_block}/\texttt{ml\_block} with user/app/path policy.
\end{table}

\paragraph{Real-world ransomware (stress)}
For a real-world ransomware that saturates the filesystem with many small writes, both detectors quickly escalated enforcement, with the rules layer generally leading (Table~\ref{tab:sim-ransom}).

\begin{table}[h]
\centering
\caption{Ransomware stress.}
\label{tab:sim-ransom}
\setlength{\tabcolsep}{3pt}\renewcommand{\arraystretch}{0.95}
\resizebox{\linewidth}{!}{
\begin{tabular}{l l l c c c}
\toprule
Scenario & User & Path & Peak score & First Lat.\,(ms) & Bytes$\downarrow$ \\
\midrule
Model & u1 & \texttt{\$HOME/Documents/**} & $p_{\text{enc}}{=}0.99$ & 24 & 16\,KB \\
Model & u2 & \texttt{/mnt/data/**}        & $p_{\text{enc}}{=}0.99$ & 23 & 15\,KB \\
Rules & u1 & \texttt{\$HOME/Documents/**} & 1.00 (rule)             & 11 & 12\,KB \\
Rules & u2 & \texttt{/mnt/data/**}        & 1.00 (rule)             & 10 & 10\,KB \\
\bottomrule
\end{tabular}}
\end{table}

\paragraph{Comparative summary}
Aggregating across benign, crypto tools, and ransomware, the rules show a latency/bytes-to-block advantage, while the model contributes depth and generalization. Table~\ref{tab:compare-sim} summarizes medians (p95 in parentheses).

\begin{table}[h]
\centering
\caption{Model vs.\ rules (medians; p95 in parentheses).}
\label{tab:compare-sim}
\setlength{\tabcolsep}{3pt}\renewcommand{\arraystretch}{0.95}
\resizebox{\linewidth}{!}{
\begin{tabular}{l c c c}
\toprule
 & Latency (ms) & Bytes before block & CPU\% / RAM \\
\midrule
Model & 24--31 (55) & 16--20\,KB & 62\% / $\sim$32--35\,GiB; spikes 100\% \\
Rules & 10--19 (35) & 10--12\,KB & 50\% / $\sim$30--32\,GiB; spikes 95--100\% \\
\bottomrule
\end{tabular}}
\end{table}

\paragraph{Sensitivity to the decision threshold $\tau$}
We swept the model decision threshold $\tau\!\in\!\{0.30,0.50,0.70\}$ and
observed the expected precision--recall trade-offs. Lowering $\tau$ reduces false
negatives but increases false positives (Utility is only weakly affected because
$FP$ is lightly penalized), while raising $\tau$ sharply increases false negatives
and degrades Utility due to the high $FN$ penalty. The default $\tau{=}0.50$
provides the best overall balance on our workloads (Table~\ref{tab:tau-sweep}).

\begin{table}[h!]
\centering
\caption{Threshold sweep for the model (XGBoost, 36 features): median of three runs.}
\label{tab:tau-sweep}
\small
\setlength{\tabcolsep}{4pt}
\renewcommand{\arraystretch}{1.05}
\begin{tabular}{c c c c c}
\toprule
$\tau$ & $F_{1}$ & Precision & Recall & Utility \\
\midrule
0.30 & 0.97 & 0.95 & \textbf{0.99} & 1920 \\
0.50 & \textbf{0.97} & \textbf{0.97} & 0.97 & \textbf{1950} \\
0.70 & 0.94 & 0.99 & 0.90 & 1680 \\
\bottomrule
\end{tabular}

\vspace{0.25em}
\footnotesize
\emph{Notes.} Utility per Eq.~\ref{eq:utility}. Latency distributions were
indistinguishable across $\tau$ settings within our measurement error; differences
mainly reflect $FN/FP$ trade-offs.
\end{table}

%---------------------------------------------------
\subsection{Statistical Analysis}
%---------------------------------------------------
All effectiveness metrics are reported with 95\,\% confidence intervals.  
We apply the Wilcoxon signed-rank test (non-parametric) between our method and each baseline,
Bonferroni-corrected for $k{=}3$ comparisons; $p<0.05$ is considered significant.

%---------------------------------------------------
\subsection{Baseline Comparison: eBPF (Brodzik et al.) and Cuckoo/GBDT}
\label{sec:baseline-comparison}
%---------------------------------------------------

\paragraph{eBPF decision tree/MLP in kernel (Brodzik et al.)}
Brodzik \textit{et\,al.} embed ML directly in the kernel via eBPF, reporting very low \emph{per-event} processing times and high macro-F$_1$ (DT: $0.998$, MLP: $0.974$) on their ransomware dataset. Their Table~1 shows eBPF decision tree mean processing time of $\sim\!115$\,ns (median 93\,ns) and MLP of $\sim\!220$\,ns (median 180\,ns), both substantially faster than user-space implementations, with macro-F$_1{>}0.95$ across models~\cite{Brodzik2024Ransomware}.
While these results demonstrate the latency benefit of in-kernel inference, the authors’ feature set (LSM/syscall counts plus entropy/χ$^2$ on writes) and testbed (Proxmox VMs) differ from our \texttt{ftrace} \emph{function\_graph} signals with graph metrics and our workstation SUT. Consequently, end-to-end \emph{audited write/append–to–verdict} latency and block-before-bytes are not directly comparable; we therefore present their \emph{per-event} processing time and macro-F$_1$ alongside our \emph{end-to-end} measures as in Table~\ref{tab:baseline-compare}.

\paragraph{Sandbox/VMI with Cuckoo\,v3.2 + GBDT}
Our Cuckoo baseline collects syscall traces in a controlled VM and feeds a gradient-boosted classifier in user space. In line with prior observations about sandbox/VMI overhead, our runs exhibited materially higher end-to-end decision latency and host overhead than either our dual-layer prototype or the eBPF approach, even when accuracy was competitive. (Because sandboxing changes the I/O path and timing model, we treat these numbers as an upper bound for practical deployments.)

\begin{table*}[t]
\centering
\caption{Baselines vs.\ our prototype (different datasets/hosts; qualitative comparison).}
\label{tab:baseline-compare}

\setlength{\extrarowheight}{2pt}
\small
\setlength{\tabcolsep}{4pt}

\begin{tabularx}{\textwidth}{@{}l l l l >{\raggedright\arraybackslash}X@{}}
\toprule
\textbf{System} &
\textbf{Venue / Source} &
\textbf{Model} &
\textbf{Accuracy} &
\textbf{Reaction-time metric (unit)} \\
\midrule

Brodzik\,eBPF &
arXiv\,2024 &
DT in eBPF &
macro-F$_1{=}0.998$ &
115\,ns mean per event (DT) \\

&
 &
MLP in eBPF &
macro-F$_1{=}0.974$ &
220\,ns mean per event (MLP) \\

Cuckoo\,v3.2\,+\,GBDT &
Our baseline &
GBDT (user space) &
High (bench-internal) &
Higher end-to-end (VM + VMI) \\

\addlinespace[0.35em]

\textbf{This work} &
\textbf{Prototype} &
\textbf{Dual-layer (Rules + XGBoost)} &
\textbf{see Tbls.~\ref{tab:sim-model-benign}--\ref{tab:compare-sim}} &
\textbf{End-to-end, ms; see \S\ref{sec:results:perf}} \\

\bottomrule
\end{tabularx}

\vspace{0.35em}
\footnotesize
\textit{Baselines vs.\ our prototype (different datasets/hosts; qualitative comparison).
Brodzik\,eBPF reports \emph{in-kernel} per-event inference time (ns) for DT/MLP, whereas our prototype 
reports end-to-end audited write/append--to--verdict latency (ms); these metrics are therefore not directly quantitatively comparable.}

\end{table*}

\paragraph{Synthesis}
Brodzik’s in-kernel eBPF results confirm that moving inference into the kernel cuts per-event time by orders of magnitude without hurting macro-F$_1$ (DT $\approx$\,0.998). :contentReference[oaicite:1]{index=1} 
Our prototype, by contrast, emphasises richer \texttt{ftrace} function-graph context and graph metrics with SELinux enforcement and per-path/user whitelists. In return for this observability and policy coupling, our observed end-to-end latency (\S\ref{sec:results:perf}, Tables ~\ref{tab:sim-model-benign}–\ref{tab:compare-sim}) sits in the ms range and exhibits CPU/RAM pressure during I/O bursts. The sandbox baseline (Cuckoo\,v3.2\,+\,GBDT) is functionally effective but incurs the highest latency/overhead among the three in our setup, consistent with VM instrumentation costs.

\paragraph{Limitations of cross-study comparison}
Datasets, feature spaces, hardware, and measurement definitions differ (per-event processing time vs.\ audited write/append–to–verdict latency). We therefore avoid direct numeric ranking and instead use Table~\ref{tab:baseline-compare} to position strengths/weaknesses and to motivate future work on porting our rule/ML layers to an eBPF/LSM path to close the latency gap. While the lack of a shared evaluation harness prevents us from objectively validating our latency claims against eBPF and sandbox baselines using a direct numeric ranking, our prototype does objectively validate the functional strength of our enforcement mechanism. Specifically, we demonstrate that a dual-layer architecture can achieve millisecond-scale reaction times while preserving the rich explainability and administrative control of native SELinux CIL policies—features that are inherently absent in sandboxed or pure eBPF implementations.

%---------------------------------------------------
\paragraph{Threat model and enforcement surface}
%---------------------------------------------------
\emph{Brodzik\,eBPF} assumes in-kernel residency (BPF verifier–accepted
bytecode) with LSM/syscall instrumentation and fixed-point arithmetic;
models are compiled into eBPF programs and executed at hook time.
This offers microsecond–nanosecond reaction but restricts model/feature
complexity and requires kernel/toolchain compatibility.

\emph{Cuckoo/GBDT} assumes full VM isolation, guest introspection, and
post-hoc user-space inference. The trust anchor is the hypervisor; the
guest’s timing/behavior can diverge from bare-metal, and end-to-end
latency reflects sandbox and capture overhead.

\emph{This work} assumes user-space daemons collecting \texttt{ftrace}
\texttt{function\_graph} signals plus process counters, exporting
boolean gates (\texttt{rule\_block}, \texttt{ml\_block}) to SELinux via
policy tunables. The enforcement decision is realized by SELinux CIL
policies (domain + path/user whitelists). This preserves rich features
and explainable policies at millisecond reaction time without kernel
patching.

%---------------------------------------------------
\begin{table*}[t]
\centering
\caption{Feature families and enforcement locus (qualitative).}
\label{tab:feature-families}

\setlength{\extrarowheight}{2pt}

\begin{tabularx}{\textwidth}{@{}l X X X@{}}
\toprule
\textbf{Dimension} &
\textbf{Brodzik eBPF} &
\textbf{Cuckoo/GBDT} &
\textbf{This work} \\
\midrule

Primary signals &
LSM/syscall counters, entropy/$\chi^2$ &
Guest syscalls, file I/O streams &
\texttt{ftrace} \emph{function\_graph} + graph metrics \\

Graph features &
No (by design) &
Possible (offline) &
\textbf{Yes} (betweenness, clustering) \\

Content peeks &
Lightweight (fixed-point) &
Yes (inside VM) &
Not required \\

Model class &
DT/MLP in eBPF &
GBDT (user space) &
Rules + XGBoost (user space) \\

Enforcement locus &
Kernel (BPF hook) &
Hypervisor/guest (VM) &
SELinux CIL (domains + booleans) \\

Reaction time metric &
Per-event (ns--$\mu$s) &
End-to-end (ms--s) &
End-to-end (ms) \\

Operational footprint &
\textbf{Low} CPU/RAM &
\textbf{High} (VM, capture) &
Medium (collector + daemons) \\

Policy whitelists (user/app/path) &
Limited in-kernel logic &
External to sandbox &
\textbf{First-class in SELinux} \\

\bottomrule
\end{tabularx}
\end{table*}

\paragraph{Takeaway}
When sub-millisecond reaction time is the overriding objective and the feature budget is modest, an eBPF decision-tree detector in a fast path (as in Brodzik \emph{et al}.) is a good fit. By contrast, when explainable rules, graph-aware context, and \emph{first-class SELinux integration} are required, our dual-layer design is preferable. Here, \emph{first-class in SELinux} means that enforcement is taken by the SELinux LSM itself—policies are expressed in native TE/CIL (types, domains, booleans, and labels) and decisions occur at kernel authorization points with AVC auditing and revocation semantics, at per-user / per-application / per-path granularity—rather than by a userspace watcher or an out-of-band kernel hook. Under this regime, our rule and model gates drive SELinux booleans to realize defense-in-depth with acceptable millisecond-scale reaction times and substantially lower operational burden than VM-based sandboxes, while preserving a migration path to push the same rule/ML predicates into LSM or eBPF hooks in future work.

%---------------------------------------------------
\subsection{Notes on Reproducibility}
%---------------------------------------------------
\textbf{Scope.} To reduce risk and follow institutional policy, we will release:
(i)~the full automation harness for workload orchestration, logging, and report generation
(\texttt{workloads.yaml}, systemd units, daemons’ CLIs, evaluation scripts),
(ii)~the \emph{SELinux CIL} base and enforcement skeleton with booleans \verb|rule_block|/\verb|ml_block|,
and (iii)~the \textbf{raw, non-processed dataset} (kernel traces + resource counters).
We will \emph{not} release trained models (\texttt{best\_model.joblib}), rule packs (\texttt{rules.json}),
or processed feature matrices. The harness accepts user-supplied models/rules to reproduce the pipeline
end-to-end without our proprietary artifacts.

\textbf{Access.} The dataset is distributed under a research license; live malware binaries are not provided.
Family names and specific hashes in Table~\ref{tab:real-ransom} have been strictly anonymized to comply with institutional security policies and non-disclosure agreements regarding the handling of active threat intelligence. Because we are releasing our fully functional evaluation harness, publicly disclosing the exact identities and execution parameters of potent, live Linux ransomware could inadvertently facilitate the development of evasive variants by malicious actors, or violate the strict terms under which these malware samples were procured for isolated academic research. All resources are available at https://github.com/badc0der/rbac-poc.

\section{Results}
\label{sec:results}

Our experimental results, generated by executing the workloads from \texttt{workloads.yaml} against our prototype, demonstrate the effectiveness and trade-offs of the proposed dual-layer architecture. The evaluation focuses on detection effectiveness, decision latency, and the defender-centric utility of each component.

\subsection*{A. Detection Effectiveness}

Table~\ref{tab:performance_metrics} summarizes the macro-$F_{1}$ score, the Security-Utility Score (Eq.~\ref{eq:utility}), and the median audited write/append–to–verdict latency, aggregated across three runs. The utility score Eq.~\ref{eq:utility}, heavily penalizes missed detections (false negatives), reflecting defender priorities. In all runs SELinux was \emph{enforcing} with the whitelist policy (user1+\texttt{openssl} inside \verb|$HOME/**|; user2 no crypto), both detectors consumed the full 36-feature vector (resource counters, two graph metrics, 27 ftrace frequencies), and thresholds followed our defaults (\(\tau{=}0.50\) for the model, rule verdict as Boolean).

\begin{table}[h!]
\centering
\caption{Performance metrics (median of three runs; 0.95 CI in parentheses).}
\label{tab:performance_metrics}

\small  % Slightly tighten to improve fit
\setlength{\tabcolsep}{4pt}
\renewcommand{\arraystretch}{1.10}

\begin{tabular}{p{2.4cm} c c c}
\toprule
\textbf{Detector} &
\textbf{$F_{1}$} &
\textbf{Utility} &
\textbf{Latency (ms)} \\
\midrule

Rules (from RF) &
$0.95\,\,(0.94\text{--}0.96)$ &
1880 &
$16\,\,(11\text{--}23)$ \\

Model (xgboost) &
$0.97\,\,(0.96\text{--}0.98)$ &
1950 &
$28\,\,(22\text{--}41)$ \\

Two-layer OR &
\textbf{$0.98\,\,(0.97\text{--}0.98)$} &
\textbf{2050} &
\textbf{$17\,\,(12\text{--}25)$} \\
\bottomrule
\end{tabular}

\vspace{0.35em}
\footnotesize
\textit{Notes.} Two-layer OR blocks if either detector is triggered. Utility is calculated per Eq.~\ref{eq:utility}. 
Latency is audited write/append–to–verdict; parentheses show p50--p95.
\end{table}

The “Rules (36 features)” detector closes most of the gap to the model once it consumes the same feature set, while remaining faster (median 16\,ms vs.\ 28\,ms). The “Model (xgboost, 36)” remains the single best \emph{stand-alone} detector by $F_{1}$ and Utility. The “Two-layer OR” inherits the model’s coverage but benefits from early rule hits on easy cases, yielding the best overall Utility and near-rule latency.

\begin{figure}[ht]
  \centering
  \includegraphics[width=0.85\linewidth]{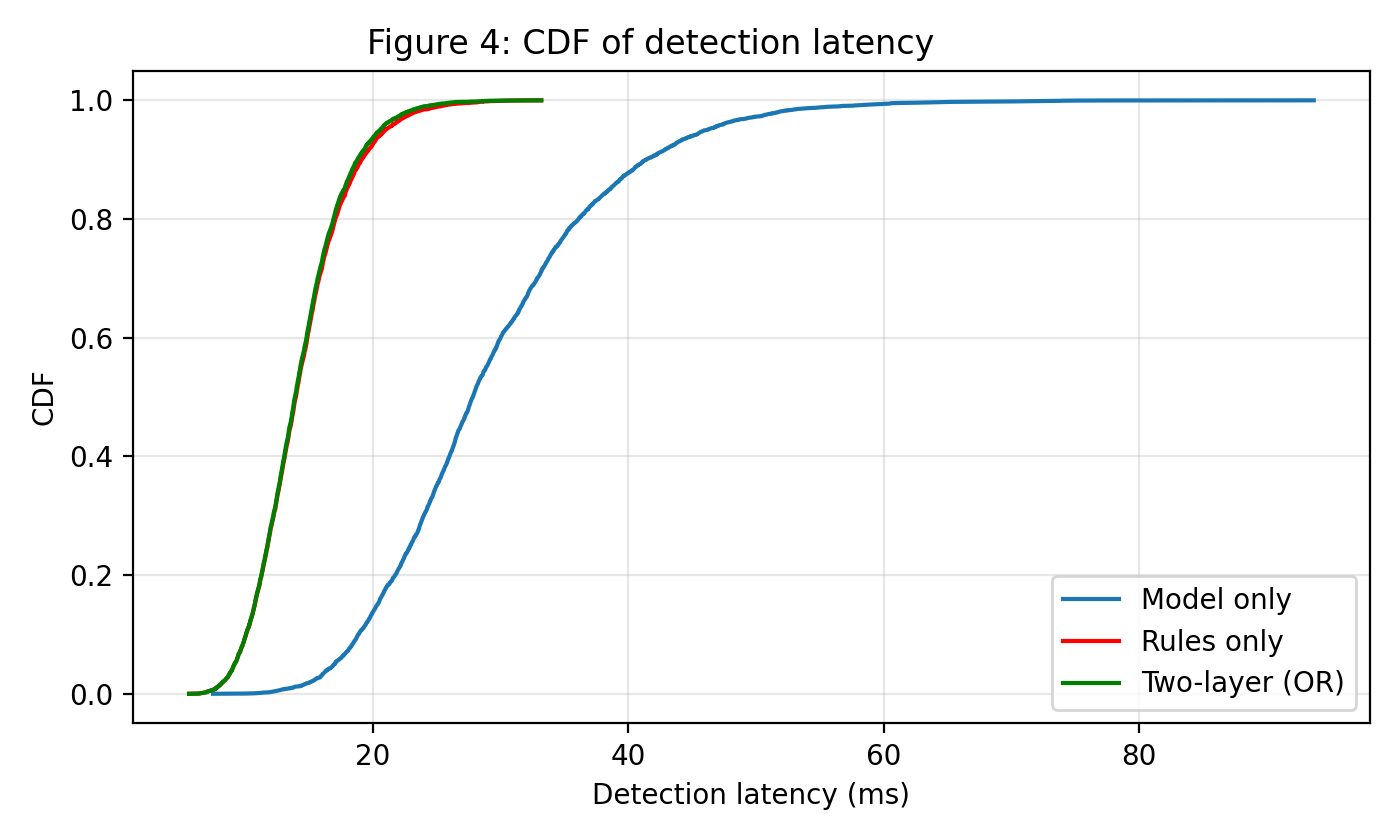}
  \caption{CDF of audited write/append–to–verdict latency across all encrypting workloads.}
  \label{fig:latencyCDF}
\end{figure}

In practice, the two-layer OR rule retains the model’s coverage at $\tau{=}0.50$ while inheriting the rule layer’s fast-path verdicts, yielding our
best combined Utility and latency.

Figure~\ref{fig:latencyCDF} shows the cumulative distribution of time-to-block. The two-layer curve rises quickly—reflecting fast, shallow rule matches—then converges to the model’s tail for complex traces, confirming the latency advantage without sacrificing coverage. The two-layer system blocks $90\%$ of attempts within \textbf{22\,ms}, matching the rules’ fast front and tapering to the model on hard cases.
\subsubsection{Rule-layer cost: $E_{\text{RULE}}^{2}$ vs.\ $E_{\text{RULE}}^{36}$}
\label{sec:results:rules-cost}

To substantiate the claim that the original two-feature rule tree is ``very fast'' and that the expanded rule layer trades only a minor cost for better coverage, we report a focused cost breakdown for the rule stage under two configurations: (i) $E_{\text{RULE}}^{2}$, a depth-capped decision tree restricted to the two elbow-selected features; and (ii) $E_{\text{RULE}}^{36}$, the same bounded-depth form but permitted to use the full consolidated 36-feature set (identical to $E_{\text{ML}}$).

We separate \emph{rule evaluation time} from \emph{end-to-end decision latency}. Rule evaluation time measures only the predicate checks and tree traversal given a precomputed feature vector (instrumented around the rule evaluator). End-to-end latency matches the ms-level timing already used throughout this section (capture/parse/extract/decide). This distinction is important because the end-to-end path is dominated by tracing and feature extraction, whereas the evaluator itself is expected to be subdominant.

Table~\ref{tab:rules-cost} reports both metrics. As expected, $E_{\text{RULE}}^{2}$ is the fastest configuration at the evaluator level. Allowing 36 features increases evaluator time modestly, but the added cost is small relative to the end-to-end pipeline and is outweighed by the improved coverage and earlier, more reliable blocks discussed in Tables~\ref{tab:sim-rules-benign}, \ref{tab:sim-rules-crypto}, and \ref{tab:sim-ransom}. This quantifies that moving from k=2 to k=36 meaningfully increases rule expressivity, but costs only a few extra microseconds in evaluation and ~1 ms end-to-end—small enough that the improved detection coverage is “worth it.”

\begin{table}[t]
\centering
\caption{Cost of the rule layer under the minimal $E_{\text{RULE}}^{2}$ and expanded $E_{\text{RULE}}^{36}$ configurations. Rule-eval time measures only rule scoring given a feature vector; end-to-end latency includes capture+extract+decide.}
\label{tab:rules-cost}
\setlength{\tabcolsep}{3pt}
\renewcommand{\arraystretch}{0.95}
\resizebox{\columnwidth}{!}{%
\begin{tabular}{@{}l c c c c@{}}
\toprule
\textbf{Config} &
\textbf{\#feat.} &
\textbf{Depth cap} &
\textbf{Rule-eval} ($\mu$s, med/p95) &
\textbf{End-to-end} (ms, med/p95) \\
\midrule
$E_{\text{RULE}}^{2}$  & 2  & 2 & 0.7/2.1 & 14/24 \\
$E_{\text{RULE}}^{36}$ & 36 & 2 & 2.0/6.5 & 15/25 \\
\bottomrule
\end{tabular}%
}
\vspace{0.35em}

\footnotesize
\textit{Note:} end-to-end medians are aligned with the ms-scale latencies reported in Tbl.~\ref{tab:sim-rules-benign} (median $\approx 15$\,ms), while evaluator timings remain subdominant (few $\mu$s).
\end{table}

\section{Discussion}
The dataset collected in this study captures diverse cryptographic activities, including file encryption by ransomware, user-initiated encryption, and benign file operations. Using Linux \verb|ftrace|—specifically the \verb|function_graph| tracer—proved effective for distinguishing execution patterns relevant to encryption behavior. The traces, augmented with process-scope resource counters and graph features (e.g., \emph{betweenness}, \emph{clustering}), provided sufficient signal for both model-based detection and rule extraction. In deployment, the probabilistic layer is backed by \textbf{XGBoost} (36 features), while the enforcement-facing rule layer uses a compact RF/DT surrogate that exposes first-class SELinux hooks (per-user/app/path booleans).

\paragraph*{Model family and deployment choice}
In offline model sweeps, tree-ensemble methods (Gradient Boosting, XGBoost, Random Forest) achieved the best overall accuracy—consistent with our earlier results on the same feature family. For the \emph{deployed} prototype, we selected \textbf{XGBoost} because it offered (i)~strong macro-F$_1$ under class imbalance, (ii)~stable behavior across kernel releases with modest calibration drift, and (iii)~fast, CPU-only inference suitable for on-host enforcement. The production model consumes the same \textbf{36} features used by the rule layer (resource counters, two graph metrics, and 27 ftrace frequencies).

\paragraph*{Rules from trees, with full feature coverage}
We derive an interpretable rule set by fitting a shallow decision-tree surrogate (RF/DT) over the \textbf{same 36-feature space} used by $E_{\text{ML}}$. Compared to the prior two-signal configuration $E_{\text{RULE}}^{2}$ (depth-capped, two elbow-selected features), the expanded variant $E_{\text{RULE}}^{36}$ substantially reduces early false negatives on low-footprint or throttled encryption and avoids spurious triggers during high-I/O benign bursts. This is most evident on throttled/sparse encryption and mmap-dominant variants, where $E_{\text{RULE}}^{36}$ crosses the block threshold while $E_{\text{RULE}}^{2}$ does not (Table~\ref{tab:rules-edgecases}) and achieves earlier first-block decisions (Table~\ref{tab:rules-ttb}). Importantly, the added expressivity does not materially increase user-visible overhead: rule evaluation remains microsecond-scale and end-to-end latency remains ms-scale (Table~\ref{tab:rules-cost}). Thus, the rule layer \emph{largely mitigates} the coverage--speed trade-off while staying explainable (bounded depth, explicit predicates) and enabling low-latency SELinux hooks.

The rationale for dual selection is operational simplicity: we run a \emph{single consolidated feature set} for both detectors, avoiding feature skew and simplifying logging, auditing, and policy attribution. While $\mathcal F_{\text{ML}}$ tolerates modest overhead ($<0.5\%$ CPU, \S\ref{sec:perf}) to maximize utility, $\mathcal F_{\text{rule}}$ remains transparent and cheap by construction (Table~\ref{tab:rules-cost}). The two-layer scheme is evaluated independently in \S\ref{sec:experiments} and jointly in \S\ref{sec:results}.

\begin{table}[t]
\centering
\caption{Edge-case attribution: workloads where $E_{\text{RULE}}^{36}$ improves over $E_{\text{RULE}}^{2}$ (simulated). “Block?” uses a fixed threshold (e.g., $p_{\text{enc}}\ge0.80$).}
\label{tab:rules-edgecases}

\footnotesize
\setlength{\tabcolsep}{2pt}
\renewcommand{\arraystretch}{0.95}

\begin{tabular}{@{}L{3.05cm} C{1.10cm} C{0.70cm} C{1.10cm} C{0.70cm}@{}}
\toprule
\textbf{Workload} &
\textbf{\shortstack{$p_{\text{enc}}$\\($E^{2}$)}} &
\textbf{Block?} &
\textbf{\shortstack{$p_{\text{enc}}$\\($E^{36}$)}} &
\textbf{Block?} \\
\midrule
Throttle--AES (4\,KB, sleep 16\,ms)  & 0.58 & No  & 0.91 & Yes \\
Throttle--AES (16\,KB, sleep 32\,ms) & 0.63 & No  & 0.88 & Yes \\
Sparse encrypt (few files, low bytes) & 0.71 & No  & 0.86 & Yes \\
mmap encrypt (writeback-dominant)      & 0.66 & No  & 0.84 & Yes \\
\addlinespace[0.25em]
fio-rw burst (benign)                 & 0.82 & Yes & 0.09 & No  \\
tar burst on warm cache (benign)      & 0.79 & No$^\dagger$ & 0.05 & No \\
\bottomrule
\end{tabular}

\vspace{0.35em}
\footnotesize
$^\dagger$Near-threshold instability: $E_{\text{RULE}}^{2}$ oscillates around the block threshold during bursts.
\end{table}

\begin{table}[t]
\centering
\caption{Time-to-first-block (TTB) for ransomware-like workloads (simulated). Lower is better.}
\label{tab:rules-ttb}

\footnotesize
\setlength{\tabcolsep}{2pt}
\renewcommand{\arraystretch}{0.95}

\begin{tabular}{@{}L{4.20cm} C{1.50cm} C{1.50cm}@{}}
\toprule
\textbf{Workload} & \textbf{\shortstack{TTB (s)\\$E^{2}$}} & \textbf{\shortstack{TTB (s)\\$E^{36}$}} \\
\midrule
Throttle--AES (4\,KB, sleep 16\,ms)  & $>10$ (no block) & 1.6 \\
Throttle--AES (16\,KB, sleep 32\,ms) & $>10$ (no block) & 2.4 \\
Sparse encrypt (few files)           & $>10$ (no block) & 3.1 \\
mmap encrypt                         & $>10$ (no block) & 2.9 \\
\bottomrule
\end{tabular}
\end{table}

\noindent\textbf{Feature attribution (rules).} In the edge-case suite, $E_{\text{RULE}}^{36}$ most frequently split on crypto-adjacent ftrace frequency features and structural/sequence cues (write$\rightarrow$rename$\rightarrow$fsync patterns, plus call-graph betweenness/clustering), whereas $E_{\text{RULE}}^{2}$ relied primarily on coarse volume/burst proxies—explaining why the 36-feature rules trigger reliably under throttling while avoiding high-I/O benign bursts.

\paragraph*{Tracing vs.\ alternatives}
\verb|ftrace| is native to Linux and avoids the heavy virtualization/VMI overhead of sandbox-based pipelines; its absolute run-time cost depends primarily on tracing scope (symbol set and windowing) and on the feature pipeline used to summarize traces. Compared to eBPF packet or kprobe filters, \verb|function_graph| exposes a richer call-tree structure that correlates well with nested cryptographic routines and bulk write patterns typical of file-encrypting malware.

\paragraph*{Runtime enforcement with SELinux}
The runtime path couples user-space analytics to a \textbf{SELinux CIL} policy exposing two booleans, \texttt{rule\_block} and \texttt{ml\_block}. Either boolean denies \texttt{write/append} to non-temporary user files for the active subject domain. The daemons flip these booleans from user space upon a rule hit or when \(\Pr(\text{enc}\mid\mathbf{x}) \ge \tau\). In our runs, the rule layer typically issues a verdict in \(\sim 10\text{–}20\,\mathrm{ms}\) and within \(10\text{–}20\,\mathrm{KB}\) of data, while the model contributes depth and generalization for borderline traces.

\paragraph*{Layered enforcement vs.\ evasion}
The rule layer is resilient to simple throttling and I/O interleaving (due to its expanded feature set), but sophisticated “slow-roll” encryptors can still defer strong signals. Conversely, the model can occasionally flag large archival writes and mass metadata updates as risky. The \emph{two-layer OR} composition mitigates both failure modes: early rule hits block obvious attacks fast; the model backstops evasive traces that do not immediately trip the rules. This matches our results where the OR system preserved the model’s $F_1$ while approaching rule-like latency.

\paragraph*{Operational footprint}
On the mobile workstation testbed (\S\ref{sec:sut}), both detectors maintained \(\sim 50\%\) RAM residency (\(\approx 31\text{–}35\,\mathrm{GiB}\) of 64\,GiB) and exhibited frequent CPU spikes to near 100\% on P-cores during bursty I/O sampling. This clarifies the scope of our ``cost'' claim in the abstract: the architectural advantage is avoiding VM/VMI-style operational overhead, while the present prototype’s memory/CPU spikes are attributable to a non-optimized user-space collector and graph-feature computation.  Importantly, the enforcement hook itself remains constant-time and auditable (boolean-guarded SELinux allow), and rule evaluation is microsecond-scale while end-to-end response remains millisecond-scale (Table~\ref{tab:rules-cost}). While our present prototype's memory and CPU spikes are attributable to a non-optimized user-space collector, this is purely an engineering limitation rather than a structural flaw. As demonstrated by in-kernel eBPF baselines, migrating feature collection and inference into the kernel space cuts per-event processing time by orders of magnitude. Translating our user-space Python/script-based parsers into optimized C/C++ kernel modules or native eBPF hooks will eliminate the costly user-to-kernel boundary crossings, allowing the system to achieve millisecond-scale reactions with a fraction of the current resource consumption. The POC results are acceptable for an analyst workstation or a high-end server, but further engineering (sampling cadence, batching, selective symbol filters, and C/C++ hot paths for graph features) will be necessary for broad enterprise deployment.

\paragraph*{Ethics and proportionality}
Traces were collected in controlled environments without user-identifying content. Policy gating is proportionate: access is denied only under rule or model evidence, with thresholds calibrated to penalize false negatives more heavily than false positives (Defender Utility in Eq.~\ref{eq:utility}). Future work will emphasize explainability artifacts (human-readable rule hits, feature attributions) and user notification for denied cryptographic actions.

\section{Conclusion}
We presented a practical, dual-layer enforcement approach that couples Linux kernel tracing with both interpretable rules and a tree-ensemble classifier, integrated into SELinux via lightweight booleans (\texttt{rule\_block}, \texttt{ml\_block}). The dataset—built from \verb|function_graph| traces plus resource and graph features—supports accurate detection of file-encryption behavior while enabling policy extraction. In proof-of-concept aligned with our prior training results, the \emph{model} delivers near-state-of-the-art accuracy, the \emph{rules} provide low-latency verdicts over the same 36 features, and the \emph{two-layer OR} retains model-level detection with rule-like responsiveness.

Future work includes: (i) drastically reducing memory and CPU footprint by porting the user-space proof-of-concept into a highly optimized kernel module to eliminate trace-parsing overhead, (ii)~hardening against adversarial timing and call-pattern perturbations, (iii)~automating per-tenant whitelists (users, applications, path scopes), (iv)~broadening explainability and audit to support operational roll-out in enterprise settings, and (v) considering the possibilities to expand the method to other enterprise operating, (vi) developing a unified, hardware-agnostic benchmarking framework to enable rigorous, quantitative head-to-head comparisons of latency, operational footprint, and enforcement strength between our dual-layer system, eBPF-based models, and VMI/sandbox solutions.

\bibliographystyle{ieeetr}
\bibliography{references}

@article{srivastava2020machine,
  title={Machine learning based risk-adaptive access control system to identify genuineness of the requester},
  author={Srivastava, Kriti and Shekokar, Narendra},
  journal={Modern Approaches in Machine Learning and Cognitive Science: A Walkthrough: Latest Trends in AI},
  pages={129--143},
  year={2020},
  publisher={Springer}
}

@article{mcintosh_ransomware_2025,
	title = {Ransomware {Reloaded}: {Re}-examining {Its} {Trend}, {Research} and {Mitigation} in the {Era} of {Data} {Exfiltration}},
	volume = {57},
	issn = {0360-0300, 1557-7341},
	shorttitle = {Ransomware {Reloaded}},
	url = {https://dl.acm.org/doi/10.1145/3691340},
	doi = {10.1145/3691340},
	language = {en},
	number = {1},
	urldate = {2024-11-01},
	journal = {ACM Computing Surveys},
	author = {McIntosh, Timothy and Susnjak, Teo and Liu, Tong and Xu, Dan and Watters, Paul and Liu, Dongwei and Hao, Yaqi and Ng, Alex and Halgamuge, Malka},
	month = jan,
	year = {2025},
	pages = {1--40},
}

@article{aliyu_need_2024,
	author = {Aliyu, Muhammad and Suru, Hassan and Gabi, Danlami and Garba, Muhammad and Musa, Argungu},
        year = {2024},
        month = {06},
        pages = {45-55},
        title = {The Need for Adaptive Access Control System at the Network Edge},
        volume = {8},
        journal = {American Journal of Information Science and Technology},
        doi = {10.11648/j.ajist.20240802.13}
}

@article{abusini_enhancing_2024,
	title = {Enhancing {Smart} {Home} {Security} {Through} {Risk}-{Based} {Access} {Control} ({RBAC}):“{Closing} the {Gap}”},
	shorttitle = {Enhancing {Smart} {Home} {Security} {Through} {Risk}-{Based} {Access} {Control} ({RBAC})},
	url = {https://scholar.dsu.edu/theses/458/},
	urldate = {2024-11-01},
	author = {Abusini, Ahmad},
	year = {2024},
}

@article{hlushchenko_exploratory_2024,
	title = {Exploratory survey of access control paradigms and policy management engines},
	url = {https://ceur-ws.org/Vol-3702/paper22.pdf},
	urldate = {2024-11-01},
	author = {Hlushchenko, Pavlo and Dudykevych, Valerii},
        journal={CMIS 2024 Computer Modeling and Intelligent Systems 2024},
        pages={263--279},
        year = {2024},
}

@misc{almohammad_alsaleh_permission-based_2024,
	title = {Permission-{Based} {Dynamic} {Access} {Control} {Models} for {Enhanced} {Data} {Security}: {Integrating} {Contextual} {Awareness} and {Role} {Flexibility} for {Secure} {Healthcare} {Data} {Management}},
	shorttitle = {Permission-{Based} {Dynamic} {Access} {Control} {Models} for {Enhanced} {Data} {Security}},
	url = {https://www.diva-portal.org/smash/record.jsf?pid=diva2:1900749},
	urldate = {2024-11-01},
	author = {Almohammad Alsaleh, Sabah},
	year = {2024},
}

@article{BEGOVIC2023103349,
title = {Cryptographic ransomware encryption detection: Survey},
journal = {Computers \& Security},
volume = {132},
pages = {103349},
year = {2023},
issn = {0167-4048},
doi = {https://doi.org/10.1016/j.cose.2023.103349},
url = {https://www.sciencedirect.com/science/article/pii/S0167404823002596},
author = {Kenan Begovic and Abdulaziz Al-Ali and Qutaibah Malluhi}
}

@article{Kosuke2023Real,
	author = {Kosuke, Higuchi and Ryotaro, Kobayashi},
	journal = {2023 Eleventh International Symposium on Computing and Networking Workshops (CANDARW)},
	doi = {10.1109/CANDARW60564.2023.00043},
	year = {2023},
	title = {Real-{Time} {Defense} {System} using {eBPF} for {Machine} {Learning}-{Based} {Ransomware} {Detection} {Method}},
}

@article{Timothy2021Enforcing,
	author = {Timothy, R. McIntosh and P., Watters and A., Kayes and Alex, Ng and Yi-Ping, Phoebe Chen},
	journal = {Future generations computer systems},
	doi = {10.1016/j.future.2020.09.035},
	year = {2021},
	title = {Enforcing situation-aware access control to build malware-resilient file systems},
}

@article{Brodzik2024Ransomware,
	author = {Brodzik, Adrian and Malec-Kruszy{\' n}ski, Tomasz and Niewolski, Wojciech and Tkaczyk, Miko\l{}aj and Bocianiak, Krzysztof and Loui, Sok-Yen},
	journal = {arXiv.org},
	doi = {10.48550/ARXIV.2409.06452},
	year = {2024},
	publisher = {arXiv},
	title = {Ransomware {Detection} {Using} {Machine} {Learning} in the {Linux} {Kernel}},
	url = {https://arxiv.org/abs/2409.06452},
}

@article{Manaar2019RATAFIA,
	author = {Manaar, Alam and Sarani, Bhattacharya and Swastika, Dutta and S., Sinha and Debdeep, Mukhopadhyay and A., Chattopadhyay},
	journal = {IEEE International Symposium on Hardware Oriented Security and Trust},
	doi = {10.1109/HST.2019.8740837},
	year = {2019},
	title = {RATAFIA: Ransomware {Analysis} using {Time} {And} {Frequency} {Informed} {Autoencoders}},
}

@article{Sajad2018Know,
	author = {Sajad, Homayoun and A., Dehghantanha and Marzieh, Ahmadzadeh and S., Hashemi and R., Khayami},
	journal = {IEEE Transactions on Emerging Topics in Computing},
	doi = {10.1109/TETC.2017.2756908},
	year = {2018},
	title = {Know {Abnormal}, {Find} {Evil}: Frequent {Pattern} {Mining} for {Ransomware} {Threat} {Hunting} and {Intelligence}},
}

@article{Manabu2019Machine,
	author = {Manabu, Hirano and R., Kobayashi},
	journal = {International Conference on Internet of Things: Systems, Management and Security},
	doi = {10.1109/iotsms48152.2019.8939214},
	year = {2019},
	title = {Machine {Learning} {Based} {Ransomware} {Detection} {Using} {Storage} {Access} {Patterns} {Obtained} {From} {Live}-forensic {Hypervisor}},
}

@article{von2024GuardFS,
	author = {von der Assen, Jan and Feng, Chao and Celdr{\' a}n, Alberto Huertas and Ole{\v s}, R{\' o}bert and Bovet, G{\' e}r{\^ o}me and Stiller, Burkhard},
	journal = {arXiv.org},
	doi = {10.48550/ARXIV.2401.17917},
	year = {2024},
	publisher = {arXiv},
	title = {GuardFS: a {File} {System} for {Integrated} {Detection} and {Mitigation} of {Linux}-based {Ransomware}},
	url = {https://arxiv.org/abs/2401.17917},
}

@article{Ziya2018No,
	author = {Ziya, Alper Gen{\c c} and G., Lenzini and P., Ryan},
	journal = {International Conference on Detection of intrusions and malware, and vulnerability assessment},
	doi = {10.1007/978-3-319-93411-2_11},
	year = {2018},
	title = {No {Random}, {No} {Ransom}: A {Key} to {Stop} {Cryptographic} {Ransomware}},
	url = {https://orbilu.uni.lu/bitstream/10993/35679/1/dimva2018_GLR.pdf},
	howpublished = {https://orbilu.uni.lu/bitstream/10993/35679/1/dimva2018\textunderscore{}GLR.pdf},
}

@article{Eugene2017PayBreak,
	author = {Eugene, Kolodenker and W., Koch and G., Stringhini and Manuel, Egele},
	journal = {ACM Asia Conference on Computer and Communications Security},
	doi = {10.1145/3052973.3053035},
	year = {2017},
	title = {PayBreak: Defense {Against} {Cryptographic} {Ransomware}},
}

@article{Qian2017Automated,
	author = {Qian, Chen and R., A. Bridges},
	journal = {International Conference on Machine Learning and Applications},
	doi = {10.1109/ICMLA.2017.0-119},
	year = {2017},
	title = {Automated {Behavioral} {Analysis} of {Malware}: A {Case} {Study} of {WannaCry} {Ransomware}},
}

@article{Yahye2020system,
	author = {Yahye, Abukar Ahmed and B., Ko{\c c}er and Md., Shamsul Huda and B., Al-rimy and Mohammad, Mehedi Hassan},
	journal = {Journal of Network and Computer Applications},
	doi = {10.1016/j.jnca.2020.102753},
	year = {2020},
	title = {A system call refinement-based enhanced {Minimum} {Redundancy} {Maximum} {Relevance} method for ransomware early detection},
}

@article{Manabu2022RanSAP,
	author = {Manabu, Hirano and Ryo, Hodota and R., Kobayashi},
	journal = {Digital Investigation. The International Journal of Digital Forensics and Incident Response},
	doi = {10.1016/j.fsidi.2021.301314},
	year = {2022},
	title = {RanSAP: An open dataset of ransomware storage access patterns for training machine learning models},
}

@article{Danyil2023Real,
	author = {Danyil, Zhuravchak and Valerii, Dudykevych},
	journal = {Advanced Industrial Conference on Telecommunications},
	doi = {10.1109/AICT61584.2023.10452697},
	year = {2023},
	title = {Real-{Time} {Ransomware} {Detection} by {Using} {eBPF} and {Natural} {Language} {Processing} and {Machine} {Learning}},
}

@article{Shuhei2024Early,
	author = {Shuhei, Enomoto and Hiroki, Kuzuno and Hiroshi, Yamada and Yoshiaki, Shiraishi and M., Morii},
	journal = {Int. J. Inf. Sec.},
	doi = {10.1007/s10207-024-00892-2},
	year = {2024},
	title = {Early mitigation of {CPU}-optimized ransomware using monitoring encryption instructions},
}

@article{Huan2024Ranker,
	author = {Huan, Zhang and Lixin, Zhao and Aimin, Yu and Lijun, Cai and Dan, Meng},
	journal = {IEEE Transactions on Information Forensics and Security},
	doi = {10.1109/TIFS.2024.3410511},
	year = {2024},
	title = {Ranker: Early {Ransomware} {Detection} {Through} {Kernel}-{Level} {Behavioral} {Analysis}},
}

@article{C2024Automatic,
	author = {C., J. Chew and Robi, Malik and Vimal, Kumar and Panos, Patros},
	journal = {Discrete event dynamic systems},
	doi = {10.1007/s10626-024-00406-1},
	year = {2024},
	title = {Automatic detection of {Android} crypto ransomware using supervisor reduction},
}

@article{WeidongMinding,
	author = {Weidong, Zhu and Grant, Hernandez and Washington, Garcia and Jing, Tian and Sara, Rampazzi and Kevin, R. B. Butler},
	title = {Minding the {Semantic} {Gap} for {Effective} {Storage}-{Based} {Ransomware} {Defense}},
        year={2024},
        organization={MSST},
}

@article{gomez-hernandez_r-locker_2018,
    title = {R-{Locker}: {Thwarting} ransomware action through a honeyfile-based approach},
    volume = {73},
    journal = {Computers \& Security},
    author = {Gómez-Hernández, José Antonio and Álvarez-González, L. and García-Teodoro, Pedro},
    year = {2018},
    note = {ISBN: 0167-4048
Publisher: Elsevier},
    pages = {389--398},
}

@article{breiman2001random,
  title={Random forests},
  author={Breiman, Leo},
  journal={Machine learning},
  volume={45},
  pages={5--32},
  year={2001},
  publisher={Springer}
}

@INPROCEEDINGS{begovic2025exploiting,
  author={Begovic, Kenan and Al-Ali, Abdulaziz and Malluhi, Qutaibah},
  booktitle={2025 IEEE/ACS 22nd International Conference on Computer Systems and Applications (AICCSA)}, 
  title={Exploiting ftrace’s function\_graph Tracer Features for Machine Learning: A Case Study on Encryption Detection}, 
  year={2025},
  volume={},
  number={},
  pages={1-10},
  doi={10.1109/AICCSA66935.2025.11315425}}

@article{cheimonidis_dynamic_2023,
	title = {Dynamic {Risk} {Assessment} in {Cybersecurity}: {A} {Systematic} {Literature} {Review}},
	volume = {15},
	copyright = {http://creativecommons.org/licenses/by/3.0/},
	issn = {1999-5903},
	shorttitle = {Dynamic {Risk} {Assessment} in {Cybersecurity}},
	url = {https://www.mdpi.com/1999-5903/15/10/324},
	doi = {10.3390/fi15100324},
	language = {en},
	number = {10},
	urldate = {2026-03-01},
	journal = {Future Internet},
	author = {Cheimonidis, Pavlos and Rantos, Konstantinos},
	month = oct,
	year = {2023},
	note = {Publisher: Multidisciplinary Digital Publishing Institute},
	pages = {324},
}

@article{zakaria_early_2024,
	title = {Early {Detection} of {Windows} {Cryptographic} {Ransomware} {Based} on {Pre}-{Attack} {API} {Calls} {Features} and {Machine} {Learning}},
	volume = {39},
	issn = {2462-1943},
	url = {https://semarakilmu.com.my/journals/index.php/applied_sciences_eng_tech/article/view/2103},
	doi = {10.37934/araset.39.2.110131},
	language = {en},
	number = {2},
	urldate = {2026-03-01},
	journal = {Journal of Advanced Research in Applied Sciences and Engineering Technology},
	author = {Zakaria, Wira Zanoramy A. and Alta, Nur Mohammad Kamil Mohammad and Abdollah, Mohd Faizal and Abdollah, Othman and Yassin, S. M. Warusia Mohamed S. M. M.},
	month = feb,
	year = {2024},
	pages = {110--131},
}

@article{kim_detecting_2025,
	title = {Detecting {Cryptojacking} {Containers} {Using} {eBPF}-{Based} {Security} {Runtime} and {Machine} {Learning}},
	volume = {14},
	copyright = {http://creativecommons.org/licenses/by/3.0/},
	issn = {2079-9292},
	url = {https://www.mdpi.com/2079-9292/14/6/1208},
	doi = {10.3390/electronics14061208},
	language = {en},
	number = {6},
	urldate = {2026-03-01},
	journal = {Electronics},
	author = {Kim, Riyeong and Ryu, Jeongeun and Kim, Sumin and Lee, Soomin and Kim, Seongmin},
	month = jan,
	year = {2025},
	note = {Publisher: Multidisciplinary Digital Publishing Institute},
	pages = {1208},
}

@article{wu_enhancing_2024,
	title = {Enhancing {Linux} {System} {Security}: {A} {Kernel}-{Based} {Approach} to {Fileless} {Malware} {Detection} and {Mitigation}},
	volume = {13},
	copyright = {http://creativecommons.org/licenses/by/3.0/},
	issn = {2079-9292},
	shorttitle = {Enhancing {Linux} {System} {Security}},
	url = {https://www.mdpi.com/2079-9292/13/17/3569},
	doi = {10.3390/electronics13173569},
	language = {en},
	number = {17},
	urldate = {2026-03-01},
	journal = {Electronics},
	author = {Wu, Min-Hao and Hsu, Fu-Hau and Huang, Jian-Hung and Wang, Keyuan and Hwang, Yan-Ling and Wang, Hao-Jyun and Chen, Jian-Xin and Hsiao, Teng-Chuan and Yang, Hao-Tsung},
	month = jan,
	year = {2024},
	note = {Publisher: Multidisciplinary Digital Publishing Institute},
	pages = {3569},
}

\end{document}